\documentclass{aa}  
\usepackage{graphicx}
\usepackage[colorlinks=true, allcolors=blue]{hyperref}
\usepackage{booktabs}
\usepackage{txfonts}
\usepackage{dirtytalk}
\usepackage{balance}

\def\sgra{\object{Sgr~A*}\xspace} 

\newcommand{\kc}{\Omega_{\mathrm{coef}}\xspace}
\newcommand{\rs}{r_{\mathrm{hs}}\xspace}

\newcommand\ipole{{\tt ipole}\xspace}
\newcommand\bipole{{\tt bipole}\xspace}

\newcommand\scipy{{\tt scipy}\xspace}

\newcommand\dynesty{{\tt dynesty}\xspace}

\begin{document} 


   \title{Lensing of hot spots in Kerr spacetime}

   \subtitle{An empirical relation for black hole spin estimation}


\titlerunning{Lensing of hot spots around black holes}

\authorrunning{Yfantis et al.}

    \author{A. I. Yfantis
          \inst{1}
          \and
          D. C. M. Palumbo\inst{2,3}
          \and
          M. Mo{\'s}cibrodzka\inst{1}
    }

    \institute{Department of Astrophysics / IMAPP, Radboud University, P.O. Box 9010, 6500 GL Nijmegen, The Netherlands\label{inst1} \\
    \email{a.yfantis@astro.ru.nl}
    \and Center for Astrophysics | Harvard \& Smithsonian, 60 Garden Street, Cambridge, MA 02138, USA \label{2}
    \and Black Hole Initiative at Harvard University, 20 Garden Street, Cambridge, MA 02138, USA\label{3}
    }

   \date{Received: April 2025; accepted YY}

 
  \abstract
  {Sagittarius~A* (\sgra) exhibits frequent flaring activity across the electromagnetic spectrum, often associated with a localized region of strong emission, known as a hot spot.} 
   {We aim to establish an empirical relationship linking key parameters of this phenomenon --emission radius, inclination, and black hole spin-- to the observed angle difference between the primary and secondary image ($\Delta PA$) that an interferometric array could resolve.}
 {Using the numerical radiative transfer code \ipole, we generated a library of more than 900 models with varying system parameters and computed the position angle difference on the sky between the primary and secondary images of the hot spot.}
{We find that the average $\Delta PA$ over a full period is insensitive to inclination. This result significantly simplifies potential spin measurements which might otherwise have large dependencies on inclination. Additionally, we derive a relation connecting spin to $\Delta PA$, given the period and emission radius of the hot spot, with an accuracy of less than $5^\circ$ in most cases. Finally, we present a mock observation to showcase the potential of this relation for spin inference.}
{Our results provide a novel approach for black hole spin measurements using high-resolution observations, such as future movies of Sgr~A* obtained with the Event Horizon Telescope, next-generation Event Horizon Telescope, and Black Hole Explorer.}


   \maketitle

\section{Introduction}
The Event Horizon Telescope (EHT) has produced images of the supermassive black holes Messier 87* (hereafter M87*) and Sagittarius A* (hereafter Sgr~A*), providing an unprecedented laboratory for testing accretion models and studying the effects of light propagation in general relativity (GR) \citep{eht:2019_paperI,eht:2022_paperI,eht_2024a}. Additionally, observations by the GRAVITY collaboration have constrained the mass of Sgr~A* to $M=4.3 \times 10^6\, M_{\odot} \,\pm 0.25\%$ \citep{Gravity2022}, establishing a crucial reference for event-horizon-scale tests of spacetime, such as black hole spin measurements and potential deviations from GR.

A major obstacle in performing precision spacetime tests from black hole images is disentangling the plasma effects of accretion from the pure GR signatures of the underlying spacetime. The combination of finite observational resolution, strong interstellar scattering along the Galactic Center line of sight, and the short dynamical timescale of Sgr~A* (approximately $30$ minutes at the innermost stable circular orbit (ISCO) for a Schwarzschild black hole) makes this separation particularly difficult. Consequently, any attempt to estimate the dimensionless spin parameter, $a_*=Jc/GM^2$, must rely on identifying a signature in one of the less plasma-dependent observables.

Among the most distinctive features of black hole spacetimes is their ability to trap photons in unstable yet bound orbits \citep{Bardeen_BH_1972}. In the case of a Kerr black hole, the shape and size of these orbits depend solely on the black hole’s mass and spin. The region where these bound orbits reside is known as the photon sphere. As photons escape these unstable orbits and reach a distant observer, they appear as a narrow ``photon ring'' on the image plane, asymptotically approaching the so-called ``critical curve'', depending on the number of windings the photon undergoes around the black hole before arrival. These features have been extensively studied in recent works, both theoretically \citep{Claude2001,JP2010,Gralla2019,Gralla2020_lensing,Prashant2024} and in connection with observations \citep{Johnson2020,Gralla2020_grtest,Wong2021}. The Black Hole Explorer (BHEX) mission \citep{Johnson_2024, BHEX2024a,BHEX2024b} has set the precise measurement of the photon ring as a key science goal.

In this work, we focus on a specific effect of BH lensing: the secondary images of localized sources, or ``hot spots'', a sub-category of photon rings. Unlike a full ring, the secondary image remains compact and forms a crescent. This provides a unique opportunity to distinguish direct from lensed photons in the image, enabling the application of the mathematical framework developed for null geodesics.  
We study two key observational effects: the time lag between primary and secondary photons ($n=0$ and $n=1$) and the difference in position angle on the screen ($\Delta PA$), with a particular focus on the latter. Specifically, we derive a new analytic approximation for calculating $\delta PA$ (where $\delta PA$ denotes the analytic estimate, while $\Delta PA$ refers to the observational measurement), propose a method to disentangle the inclination angle ($i$) from $\Delta PA$ measurements, and establish an empirical relation connecting $\Delta PA$ to BH spin. This method is entirely achromatic, as it depends solely on the geometric properties of the system rather than specific emission mechanisms. While the appearance of a single hot spot may differ between, e.g. radio and near-infrared wavelengths, the geometric relations underlying our method remain valid in both regimes. 

Observations of Sgr~A* support the presence of emission from localized regions, as the source exhibits regular flaring events across the electromagnetic spectrum on characteristic timescales of $\sim 20$ hours. During these events, near-infrared (NIR) radiation can increase by 1–2 orders of magnitude \citep{gravity:2020c}, while the brightest X-ray flares reach peaks $\sim 600$ times the quiescent state \citep{sgra_xflares2019}. Additionally, GRAVITY has detected astrometric motion of the Sgr~A* brightness centroid and polarimetric signatures associated with flares \citep{gravity:2018,G23}. At radio frequencies, observations with the Atacama Large Millimeter/submillimeter Array (ALMA) have linked similar physical effects to corresponding signatures in the light curves of linear polarization \citep{Wielgus2022_LC,W22,eht:2022_paperII}.  Proposed next-generation improvements to the EHT have specifically targeted improved dynamical imaging sensitivity on the Galactic Center due to the presence of these rapidly evolving events \citep{Doeleman_2023}.

Numerous studies have modeled this localized emission \citep[e.g.,][]{Brod-Loeb2006, Trippe2007, Hamaus2009, Tiede2020,Gelles:2021,Vos:2022,Vincent2023}, and some have sought to connect simulations to observations for model classification \citep{gravityMichi, gravityAlejandra,ball:2021,aimar_2023_plasmoid,yfantis24b,yfantis24a,Antonopoulou24}. As these models increasingly reproduce observational data, a stronger theoretical foundation becomes essential.  

The broader concept of hot spots has been linked to physical processes in magnetically arrested disks \citep[MAD;][]{Narayan2003}, which exhibit flux eruption events resembling the flaring activity of Sgr~A* \citep{dexter20,Scepi2022}. These events can generate flux tubes—evacuated regions in the accretion flow where electrons become highly accelerated and heated via magnetic reconnection while traveling in strong vertical magnetic fields \citep{porth21, Ripperda2022}. Simulations by \cite{Najafi2024} suggest that such structures may produce signatures consistent with observations.  
Another possible explanation is the formation of plasmoids via magnetic reconnection in magnetized disks \citep[e.g.,][]{Rip_2020_plasmoid,aimar_2023_plasmoid,Vos_2024_plasmoid}. Lastly, an alternative phenomenon that could appear as bright "spots" are the so-called shock waves (appearing as spiral arms) as shown in \cite{Conroy:2023kec}. Notably, all these phenomena are theorized to appear with non-Keplerian (typically sub-Keplerian) velocities.

Given the growing scientific progress in both observational and modeling efforts, we anticipate that a direct image or time-resolved movie of a secondary image will become a reality in the near future. In this paper, we present a straightforward and computationally efficient approach to interpreting such observations -- contrasting with the full analytic treatment of photon rings -- with the goal of constraining BH spin. To this end, we have constructed a library of over 900 models, varying key system parameters: the spot’s orbital radius ($r_{\textrm{hs}}$), orbital velocity, BH spin ($a_*$), and inclination angle ($i$).  

In Section~\ref{sec:math0}, we summarize the key mathematical results from previous studies. In Section~\ref{sec:num}, we present our numerical scheme and the model library we use to study the problem. In Section~\ref{sec:results}, we introduce our intermediate findings: a simple and highly accurate (sub-degree) approximation for $\delta PA_{0\rightarrow1}$, (as opposed to $n\rightarrow \infty$ that is typically considered); along with a simplified formula that has has a growing deviation with spin from sub-degree to up to $5^\circ$ for $a_*=0.99$. We then show that the mean value of $\Delta PA$ over a full orbit of the hot spot is insensitive to $i$, similar to recent results in \cite{Rahul_2024aXv}. This result favors simple analyses of average deflections, removing the notion of time-dependent orbital phase and the observer's inclination and position angle in downstream fitting.  Lastly, we introduce an empirical relation derived from our model library and evaluate its accuracy in reproducing simulation results. We then apply this relation to three sets of mock observations, incorporating reasonable uncertainties for $\Delta PA$, orbital period ($P$), and $r_{\textrm{hs}}$. We show that in the general case our method can constrain the BH spin to better than $\pm 0.3$ at $2\sigma$ confidence level. In Section \ref{sec:discussion}, we conclude by discussing the implications and potential limitations of this approach for future observations.

\section{Mathematical prior}\label{sec:math0}

In this subsection we use Boyer-Lindquist coordinates $(t, r, \theta, \phi)$ and natural units, i.e., $G=c=M=1$, so that both time and distance are measured in [M]. The line element describing the spacetime is 

\begin{equation} 
\begin{aligned}
ds^2 = -\left(1-\frac{2r}{\Sigma}\right)dt^2 - \frac{2a_*r\sin^2{\theta}}{\Sigma}dtd\phi + \frac{\Sigma}{\Delta}dr^2 + \Sigma d\theta^2\ + \\ \left(r^2+a_*^2 +\frac{2a_*^2r \sin^2{\theta}}{\Sigma} \right)\sin^2{\theta}d\phi^2 
\end{aligned}
\end{equation}
where $a_*$ is the dimensionless angular momentum per unit mass and
\begin{equation}
    \Sigma(r, \theta) = r^2 + a_*^2 \cos^2(\theta) ,\,\,\,
\Delta(r) = r^2 - 2r + a_*^2.
\end{equation}

We consider a stationary hot spot and show solutions for two quantities: the time lapse ($\delta t$) between two subsequent images (n, n+1) and the difference in their position angle on the sky ($\delta PA$) for an observer at infinity, on the symmetry axis. The solution holds only for $n\rightarrow\infty$, implying that the parameters are evaluated at $r_{\textrm{crit}}$. For the purpose of this paper we skip the complete solution that contains various steps and different cases, which can be found in numerous papers \citep[e.g.][]{Beckwith_2005,Gralla2020, Wong2021}. Instead we provide the solutions similarly to \cite{Gralla2020} as:

\begin{multline}
\label{eq:dt}
    \delta t = \frac{2}{\sqrt{\tilde{b}^2 - a_*^2}} 
\Bigg[\tilde{r}_0^2 \left( \frac{\tilde{r}_0 + 3}{\tilde{r}_0 - 1} \right) 
K\left( \frac{a_*^2}{a_*^2 - \tilde{b}^2} \right) \\
- 2a_*^2 E'\left( \frac{a_*^2}{a_*^2 - \tilde{b}^2} \right)\Bigg]
\end{multline}
and
\begin{equation}
\delta \textrm{PA} = \pi + \frac{2a_*}{\sqrt{\tilde{b}^2 - a_*^2}} 
\left( \frac{\tilde{r}_0 + 1}{\tilde{r}_0 - 1} \right) 
K\left( \frac{a_*^2}{a_*^2 - \tilde{b}^2} \right),
\label{eq:delta0}
\end{equation}
where 
\begin{equation}
\tilde{r}_0 = 1 + 2\sqrt{1 - \frac{a_*^2}{3}} 
\cos\left[
\frac{1}{3} \arccos\left(
\frac{1 - a_*^2}{\left(1 - \frac{a_*^2}{3}\right)^{3/2}}
\right)
\right], 
\end{equation}
is the critical radius, denoting the bound photon orbit for the face-on observer ($i = 0^\circ$). The tilde on the rest of the parameters indicates that they are calculated on the critical radius. Then $\tilde{b}$, the apparent critical curve on an observer's screen in units of M, is given by

\begin{equation}
\tilde{b} = \sqrt{\frac{\tilde{r}_0^3}{a_*^2} 
\left[
\frac{4 \Delta(\tilde{r}_0)}{(\tilde{r}_0 - 1)^2} - \tilde{r}_0
\right] + a_*^2}.    
\end{equation}

Lastly, $K$ and $E'$, in Equations~\ref{eq:dt} and \ref{eq:delta0}, represent the first and second complete elliptical integrals respectively, which can be found in math libraries, such as \scipy. 

For the case of a moving hot spot on a circular orbit on the equator there is an extra degree of complexity, since the spot changes position while the photons move to the secondary image, resulting in a necessary correction for $\delta\text{PA}$ (Equation~\ref{eq:delta0})

For the case of a moving hot spot there is an extra degree of complexity, since the spot changes position while the photons move to the secondary image, resulting in a necessary correction for $\delta\text{PA}$ (Equation~\ref{eq:delta0}). In the simplest case of a circular orbit on the equator it is given by, 

\begin{equation}
    \delta\textrm{PA}_{\textrm{orb}} = \delta\text{PA} - \delta t \Omega\,, 
    \label{eq:delta_om}
\end{equation}
where $\Omega$ is the angular velocity of the spot. 
 
As stated, these equations describe the problem only for higher-order images ($n\rightarrow\infty$). In this paper however we are interested in the secondary image ($n=1$), so we need to quantify how much the solution deviates in this case. To our knowledge, there is no published solution for such a case, hence we present it in Section~\ref{sec:results}.

\section{Numerical method}
\label{sec:num}

\begin{figure*}[ht]
    \centering
    \includegraphics[width=0.941\linewidth,trim={0.cm 0.0cm 0 0.cm},clip]{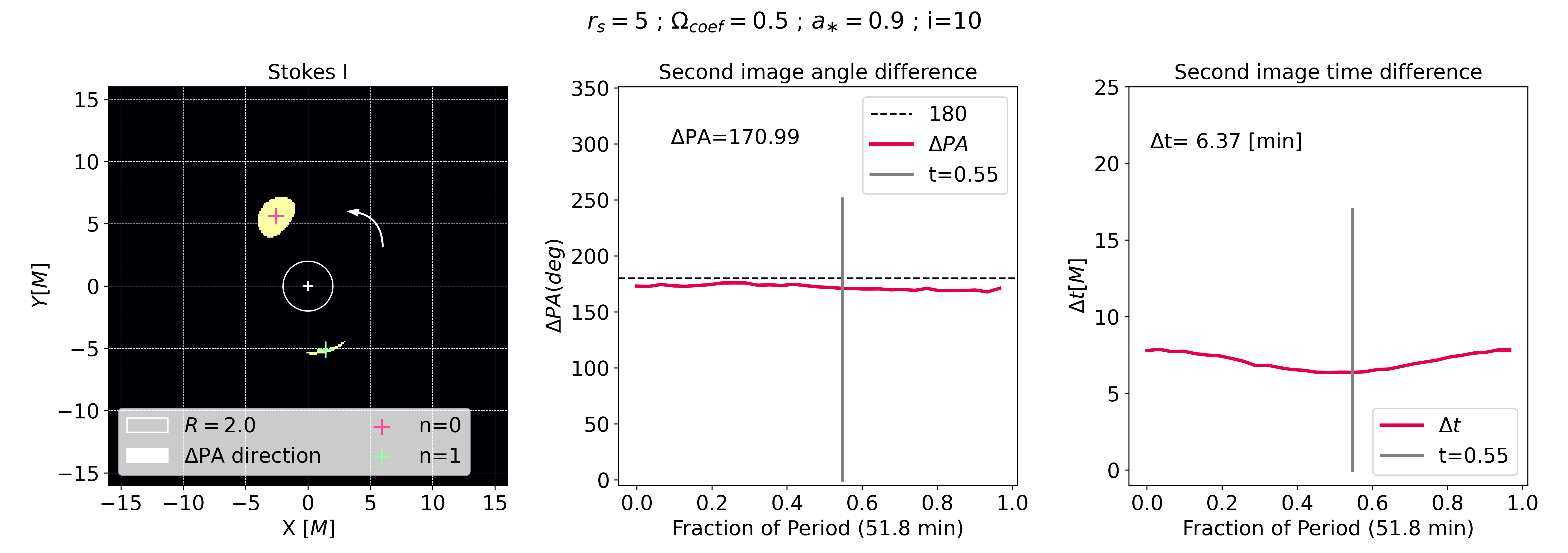} \\

        \includegraphics[width=0.941\linewidth,trim={0.2cm 0.0cm 0 0.2cm},clip]{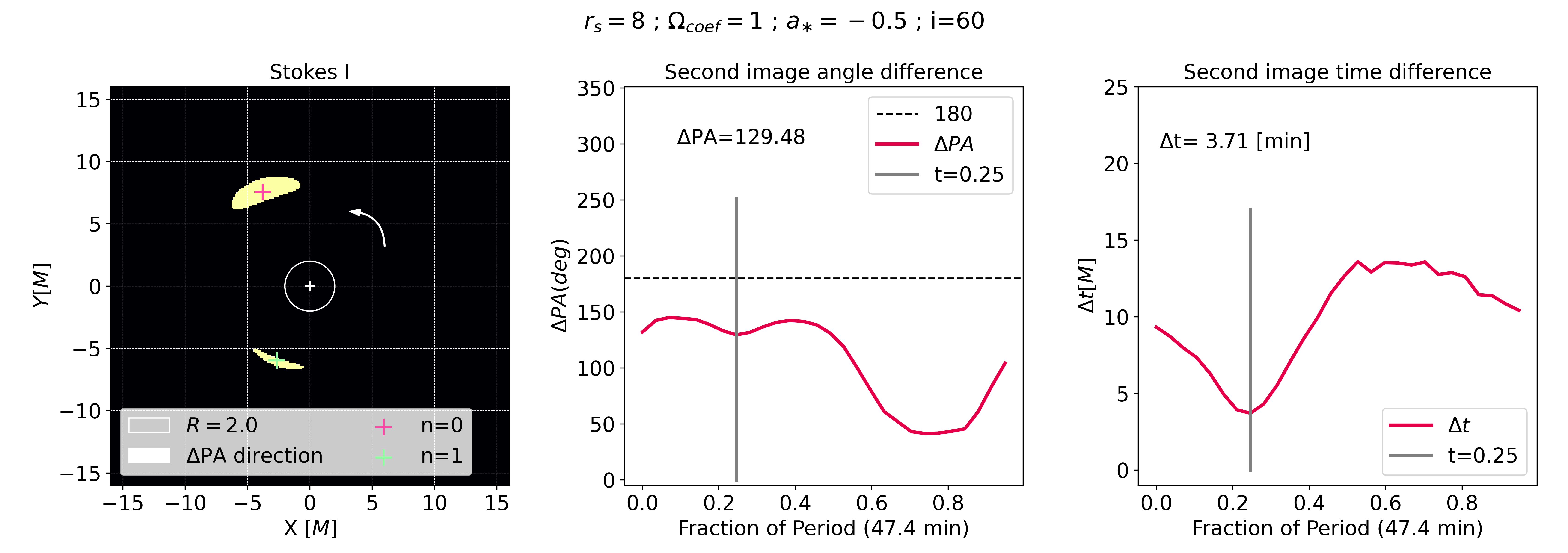}
   
    \caption{Left: Snapshot from a hot-spot movie ; middle: the curve of $\Delta PA$ measurement during a full period; right: the time lag measurement (arrival time difference) between the first and second image.}
    \label{fig:ipole_iamage}
\end{figure*}

In this section we describe our numerical setup, built inside the general relativistic radiative transfer (GRRT) code \ipole \citep{Monika2018}. We have simulated emission from a moving hot spot, closely resembling that implemented in \cite{yfantis24b,yfantis24a}. The hot spot lives in circular orbits on the equatorial plane, and it has a Gaussian profile peaking at the center for its key parameters: number density ($n_e$), electron temperature ($\Theta_e$) and magnetic field strength ($B_{\textrm{field}}$). Its size is dictated by the number density that has a $1\sigma$ radius of $1.5$M and then a steep cutoff, creating a crisp spot size. While the spot model is defined in BL coordinates, the ray-tracing of photons is done in Kerr-Schild (KS) coordinates. The resolution is set at $256\times256$ pixels. 

Our library consists of models with different values for a variety of 
parameters ($r_{\textrm{hs}},\, \kc,\, i, \, a_*$), where $r_{\textrm{hs}} $ is the distance of the spot from the center of the BH and the $\kc$ parameter controls the orbital velocity as 
\begin{equation}
    P=\frac{2\pi}{\kc}\left(r_{\textrm{hs}}^{1.5} + a_*\right). 
\label{eq:P}
\end{equation} 
The $\kc$ parameter provides a straight-forward way to simulate sub-Keplerian orbits, highly expected in MADs, or even slower orbits associated with spiral shock waves. The parameter grid is summarized in Table~\ref{tab:par}.

\begin{table}[ht]
    \centering
        \caption{ Parameter values in our simulated library.}

    \begin{tabular}{c c }
        \hline
        Parameter & Values \\
        \hline
        $r_{\textrm{hs}}$ & $\{4,5,6,7,8,9,10,11,12\}$  \\
        $\kc$ & $\{0.25,0.5,0.75,1,1.25\}$  \\
        $i$ & $\{0,10,20,25,30,60\}$  \\
        $a_*$ & $\{-0.9,-0.5,0,0.5,0.9\}$ \\
        \hline
    \end{tabular}
    \label{tab:par}
\end{table}

During the simulation we record values of the centroids of the two images, and we mask all contributions to light rays from $n>1$. Given the centroids we define $\Delta PA$ as: 
\begin{equation}
    \Delta PA= \textrm{atan2}(Y_{\textrm{cen}_{2}}X_{\textrm{cen}}-X_{\textrm{cen}_{2}}Y_{\textrm{cen}}\,,\,X_{\textrm{cen}}X_{\textrm{cen}_{2}}+Y_{\textrm{cen}}Y_{\textrm{cen}_{2}}) \, 
\end{equation}
where $X_{\textrm{cen}},Y_{\textrm{cen}}$ and $X_{\textrm{cen}_{2}},Y_{\textrm{cen}_{2}}$ refer to the centroid of total intensity of primary and secondary images respectively. It is worth mentioning that the correct $\Delta{PA}$ of a moving source can be estimated only using the slow light approach; otherwise, using the fast light approximation, the second term of Equation \ref{eq:delta_om} cannot be manifested, leading to large deviations \footnote{Note, that ray-tracing of GRMHD snapshots is typically done in fast light \citep{eht:2019e,eht:2022_paperV}.}.  

\begin{figure*}
    \centering
    \includegraphics[width=0.4841\linewidth,trim={0.2cm 0.0cm 0 0.2cm},clip]{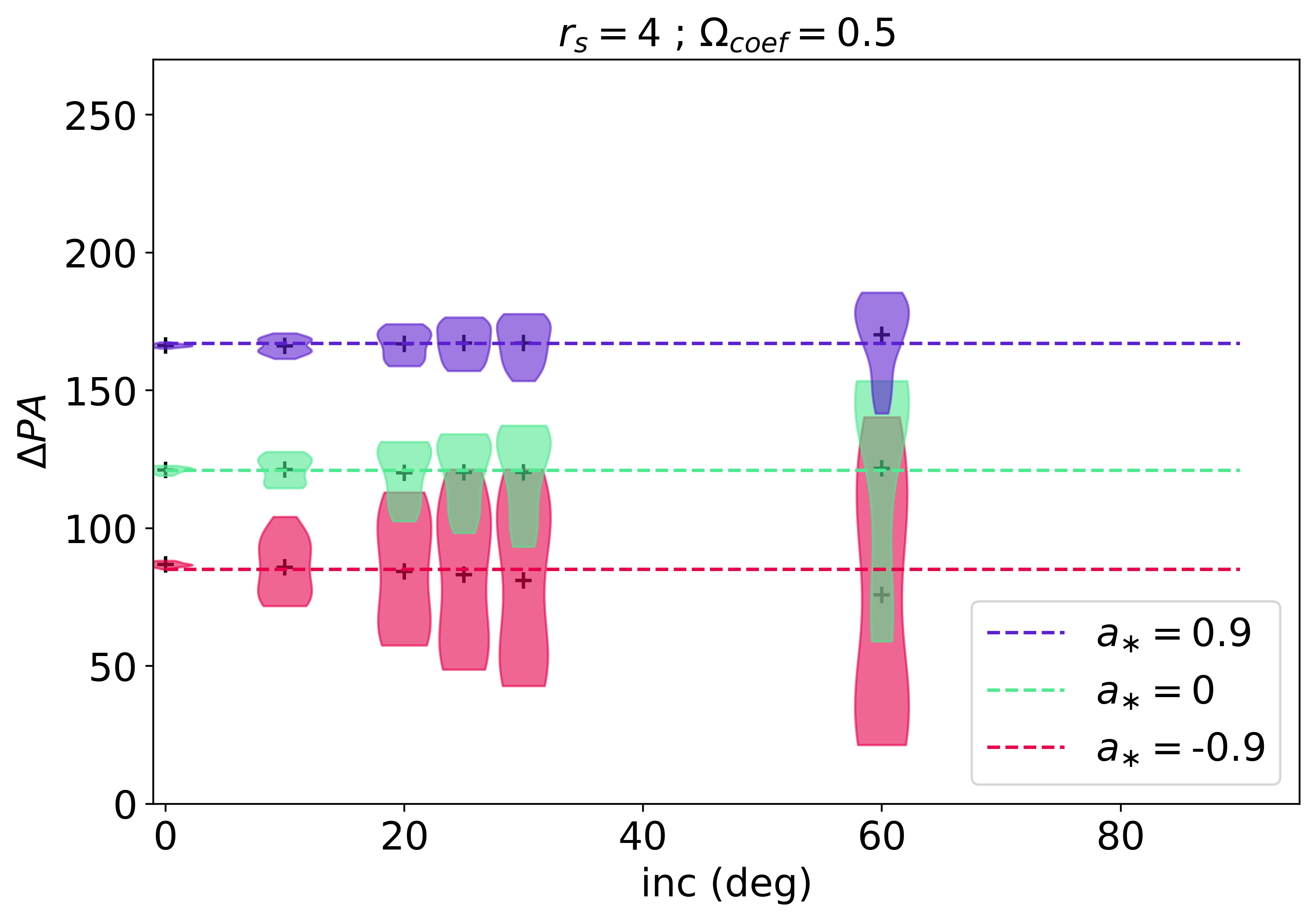}
\includegraphics[width=0.4841\linewidth,trim={0.2cm 0.0cm 0 0.2cm},clip]{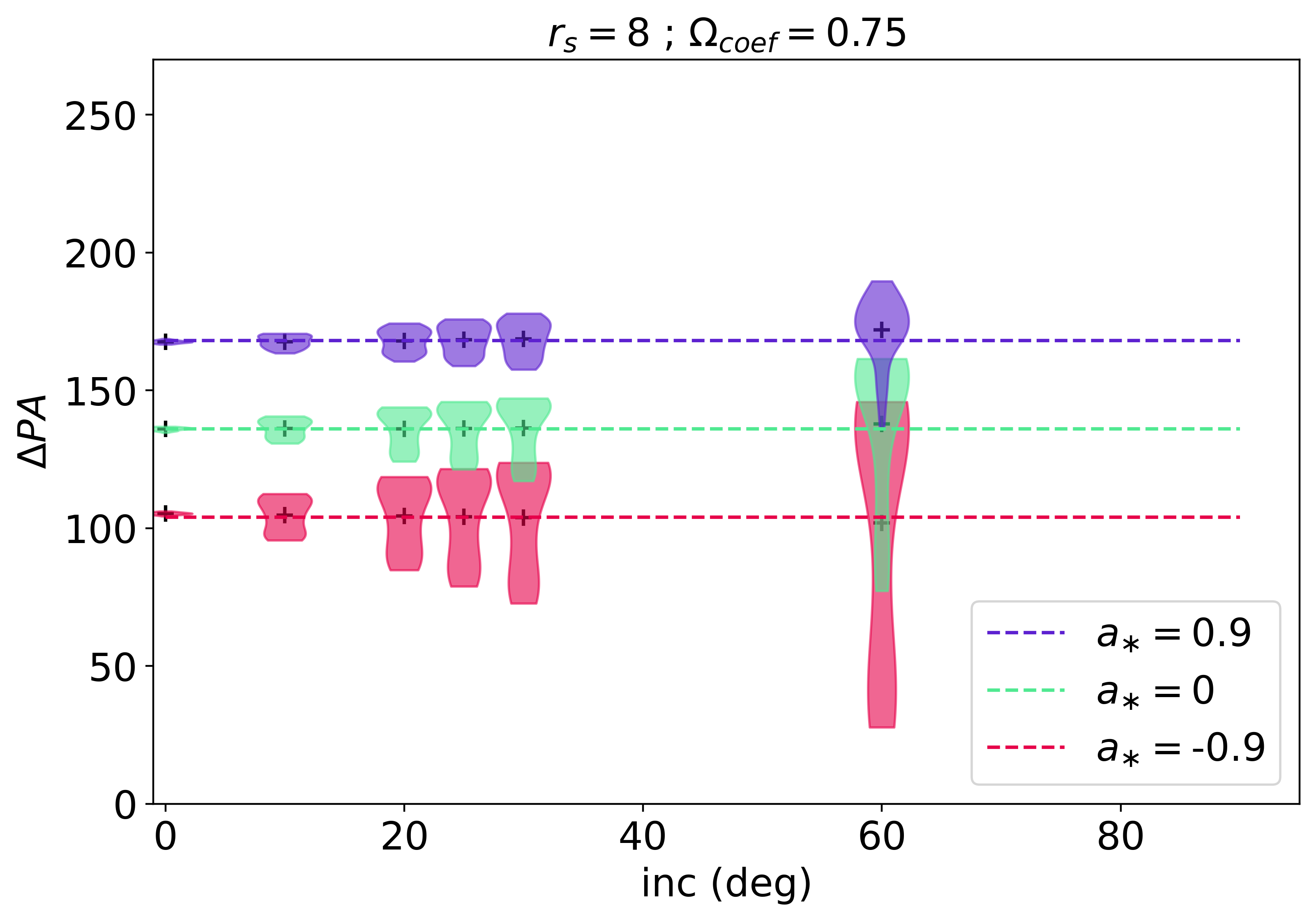}\\
    
    \caption{Distribution (violins) of $\Delta PA$ values during a full period for different inclination values. The crosses denote the average values per period and the dashed lines the value at $i=0^\circ$ for a fixed $a_*$.}
    \label{fig:dist_dpa_inc}
\end{figure*}

We also measure the arrival time of all photons to the observer, allowing us to calculate the maximum time difference ($\Delta t$) between the primary and secondary image as the difference between the first and last photon to reach. In Figure \ref{fig:ipole_iamage}, we present two example snapshots from the simulation domain with different parameters, as indicated in the plot title. The middle and right panels display the evolution of $\Delta PA$ and $\Delta t$ over a full orbit. Both quantities exhibit significant variation with increasing inclination, highlighting the need for a method to disentangle inclination effects from other influences. This is presented in the following Section~\ref{sec:results}. 

\section{Results}\label{sec:results}

This section includes two intermediate results necessary for the construction of an empirical relation. Those are: 1) an analytic approximation for $\delta PA_{0\rightarrow1}$, and 2) the independence of the mean $\Delta PA$ (from a full period) from changes in inclination, similarly to the independence of the median value shown in \cite{Rahul_2024aXv}. Table \ref{tab:notation} summarizes key parameters of the lensing problem used across the paper. Next we present our empirical relation that covers shortcomings from the analytic relations in order to be used as a means of inference, and lastly we show three examples of its application using mock observations of $\Delta PA$.

\begin{table*}[ht]
    \centering
        \caption{Descriptions of various notations used in the analysis.}
    \begin{tabular}{c l }
        \hline
        Notation & Description  \\
        \hline
        $\delta PA$ & The analytic solution for angle difference on sky between $n$, $n+1$ ($n\rightarrow\infty$), for a stationary spot. \\
        $\delta \textrm{PA}_{0\rightarrow{1}}$ & The analytic solution for angle difference on sky between $n=0$, $n=1$ for a stationary spot. \\
        $\delta \textrm{PA}_{{f2,\,}0\rightarrow{1}}$ & The analytic approximation (remove \say{factor 2}) for angle difference on sky between $n=0$, $n=1$ for a stationary spot. \\
        $\delta PA_{\textrm{orb}}$  & The analytic solution for angle difference on sky between $n$, $n+1$ ($n\rightarrow\infty$), accounting for emitter motion. \\
        $\Delta PA$ & The observed angle difference on sky between $n=0$, $n=1$, accounting for emitter motion. \\
        $\delta t$ & The analytic solution for time difference between $n$, $n+1$ ($n\rightarrow\infty$). \\
        $\Delta t$ & The observed time difference between $n=0$, $n=1$ \\
        \hline
    \end{tabular}
    \label{tab:notation}
\end{table*}

\subsection{Analytic approximation for \texorpdfstring{$\delta PA_{n\rightarrow1}$}{}}

In order to address this problem we create three models with spin parameters $a_*=0,\,0.5,\, 0.95$, a smaller spot size (spot radius of $1$ M) and higher resolution ($1024\times1024$ pixels). The spot is stationary and $r_{\textrm{hs}}$ is set to the critical radius for all models.

First we test the simulation results against the $\delta PA$ solution. While both methods give $\Delta \text{PA}=180^\circ$ for the non-spinning case, for $a_*=0.5,\, 0.95$ the analytic solution gives $\delta\text{PA} = 216^\circ, 263^\circ$ while the simulations produce $\Delta\text{PA} = 198^\circ, 218^\circ$, respectively. The reason is that the trajectories are sufficiently different for $n=0$ and $n=1$ that the critical parameters do not apply exactly. Thus the complete integral ($K$) in Equation \ref{eq:delta0} should be calculated as incomplete instead ($K_{\textrm{inc}}$). These are connected via 

\begin{equation}
    K(x) = K_{\textrm{inc}}\left(\frac{\pi}{2}\, \Bigg| \,x\right), 
\end{equation}
where the argument in $K_{\text{inc}}$ is an angle in the first part and the calculated quantity in the second. So for $\pi/2$ the incomplete becomes complete. 

Via trial and error we found that the correct angle for an incomplete integral to account for the primary-secondary ($0\rightarrow1$) image solution ($\delta \textrm{PA}_{0\rightarrow{1}}$) is $\pi/(\pi+a_*+0.3)$ so that 

\begin{multline}
\delta \textrm{PA}_{0\rightarrow{1}} = \pi + \frac{2Ma_*}{\sqrt{\tilde{b}^2 - M^2a_*^2}} 
\left( \frac{\tilde{r}_0 + M}{\tilde{r}_0 - M} \right)\times \\
K_{\textrm{inc}}\left(\frac{\pi}{(\pi+a_*+0.3)}\, \Bigg|\, \frac{a_*^2}{a_*^2 - \tilde{b}^2} \right).
\label{eq:delta01}    
\end{multline}
The reason behind the $a_*$ dependency inside $K_{\textrm{inc}}$ is that the deviation of the two formulas ($\delta\textrm{PA}$, $\delta \textrm{PA}_{0\rightarrow{1}}$) increases for high spins, as the photons are traveling a larger angle ($\delta PA$ increases with increasing spin), and that needs to be accounted for. Another effective and significantly simpler way to calculate the deviation in $\delta \textrm{PA}_{0\rightarrow{1}}$ is by taking the expression for $\delta\textrm{PA}$ (Equation~\ref{eq:delta0}) and removing the factor of 2 from the second term. We call this $\delta PA_{f2,\,0\rightarrow1}$ given as, 
\begin{multline}
\delta \textrm{PA}_{f2,\,0\rightarrow{1}} = \pi + \frac{a_*}{\sqrt{\tilde{b}^2 - a_*^2}} 
\left( \frac{\tilde{r}_0 + 1}{\tilde{r}_0 - 1} \right)
K\left( \frac{a_*^2}{a_*^2 - \tilde{b}^2} \right).
\label{eq:deltaf2}    
\end{multline}

This works nearly exactly (sub-degree deviations) for spins up to $\sim0.75$ with deviations of just a couple degrees above that ($5^\circ$ for $a_*=0.99$). A naive but perhaps useful intuition for this approximation is that the $n=0$ photons do not feel the twist of the spacetime in \say{their way in}, only as they move towards the back of the BH, and so the deflection between $n=1$ and $n=0$ reflections only half of the effect in the universal regime.

Despite these simplifications, the analytic problem remains challenging due to two key generalizations: the motion of the hot spot and the treatment of a general emission radius $r_{\textrm{hs}}$, which is not fixed at the critical radius. We therefore proceed with an extensive numerical parameter study to identify a simpler empirical relation.

\subsection{Inclination dependence}

In order to study the variations of $\Delta PA$ with inclination we create 
distributions (violins) of all $\Delta PA$ values, sampled every $dt=5M$ over a full period, for various models. These are shown in Figure \ref{fig:dist_dpa_inc}, where the models are characterized by two fixed parameter sets $(r_{\textrm{hs}}=4, 8$ and $\kc=0.5, 0.75)$, three spin values ($a_*=-0.9, 0, 0.9$), and five inclination angles in the range $[0, 60]$. The two parameter sets ($r_{\textrm{hs}}, \kc$) are shown in separate panels, while all spin values are overlaid within each panel. The vertical axis represents $\Delta PA$, and the horizontal axis denotes the inclination angle. 

The distributions shown in Figure \ref{fig:dist_dpa_inc} exhibit significant variations in both $\Delta PA$ values and their shapes depending on the parameters. However, a clear pattern emerges: the mean $\Delta PA$ for each model, marked by crosses and connected with dashed lines, remains nearly constant across inclinations. This holds especially well for inclinations up to $30^\circ$ and for larger values of $r_{\textrm{hs}}$. Even in the case of $r_{\textrm{hs}}=4$, the deviation at $i=60^\circ$ from $i=0^\circ$ is $5^\circ$. In the case of \sgra inclination is constrained by various observations to be $i\lesssim30^\circ$ \citep[e.g.][]{gravity:2018,gravityMichi,eht:2022_paperV, eht:2024b,W22,Aviad2024,yfantis24a,yfantis24b}. Given a full-period dataset, the mean $\Delta PA$ can be reliably used for inclination effects. Furthermore, the spread of $\Delta PA$ values provides a potential means for estimating inclination itself. This important property enables the creation of a simple empirical relation.

\subsection{Empirical relation for \texorpdfstring{$\Delta PA$}{}}
We find the following empirical relationship connecting three fundamental parameters of a hot spot around a BH ($r_{\textrm{hs}},\,\kc,\,a_*$) with the observed angle difference between primary and secondary image ($\Delta PA$):

\begin{equation}
    \Delta PA \, [{}^\circ] = 180^\circ + a_*(40-r_{\textrm{hs}}) - \left(\frac{1408-233a_*}{r_{\textrm{hs}}^2} +35\right) \kc \,,
\label{eq:empirical}
\end{equation}
where $\kc$, can be expressed in terms of $P,\,r_{\textrm{hs}}$ and $a_*$ by rearranging Equation \ref{eq:P} (and using physical units) as, 

\begin{equation}
    \kc=2\pi \frac{r_{\textrm{hs}}^{1.5}+a_*}{P[\text{M}]}  =2.2248\,\frac{r_{\textrm{hs}}^{1.5}+a_*}{P[\text{min}]}\,\
    \label{eq:Kc}
\end{equation}
where the last expression includes mass of \sgra from \cite{Gravity2022}, $M=4.3 \times 10^6 M_\odot$.

There are three key improvements of Equation \ref{eq:empirical} with respect to the deflection Equations~\ref{eq:delta01} and \ref{eq:deltaf2}. 
\begin{enumerate}
    \item The addition of the angular lag caused by the movement of the emitter. This is present in Equation \ref{eq:delta_om}, but the complexity increases significantly with the inclusion of $\delta t$ (Equation \ref{eq:dt}). 
    \item The flexibility of $\rs$, taking all possible values, in contrast to the Equations \ref{eq:delta01}, \ref{eq:deltaf2}, that are restricted to $\rs$ equal to the critical curve.
    \item The simplicity and computational efficiency achieved with Equation \ref{eq:empirical}, since the elliptic integrals $E,\,K$ are eliminated. 
\end{enumerate}

Regarding the individual terms of Equation \ref{eq:empirical}, the second term, $a_*(40-r_{\textrm{hs}})$, approximates the angular lag of a stationary emitter. As a single numerical demonstration, we find in a ray-traced hot spot with $a_*= 0.5$ and $r_{\textrm{hs}} =2.88$ (the critical radius for this spin) that $\Delta PA = 198.5$, near identical to $\delta PA_{0\rightarrow1}=198$. The next term includes orbital motion through a linear dependence on $\kc$, similarly to $\delta PA_{\textrm{orb}}$ (Equation \ref{eq:delta_om}). The coefficient $\kc$ is captures variation of both $r_{\textrm{hs}}$ and $a_*$. We use fitting procedures to find the correct coefficients in Equation \ref{eq:empirical}. 

\begin{figure*}
    \centering
    \includegraphics[width=0.791\linewidth,trim={0.2cm 0.0cm 0 0.2cm},clip]{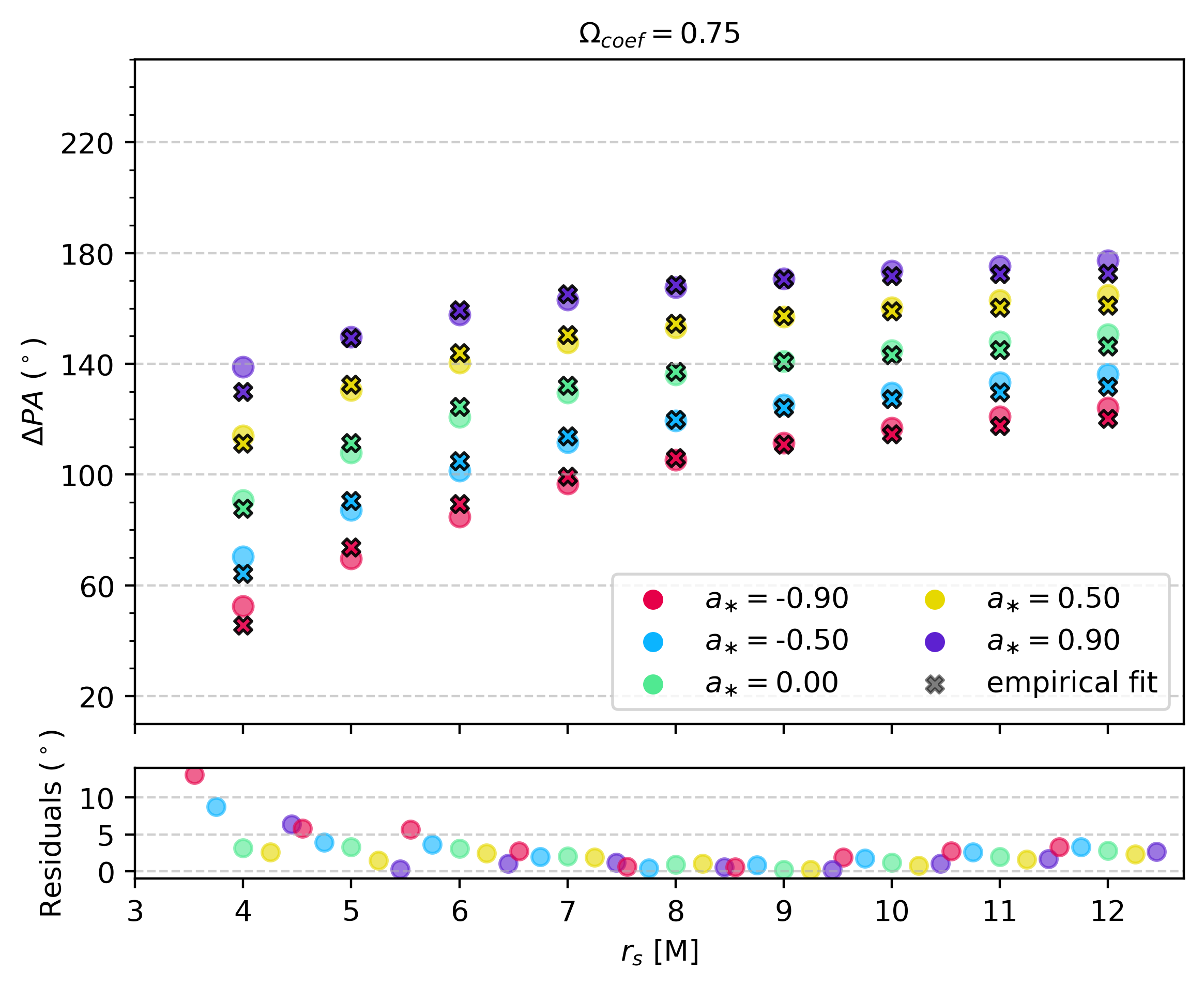}

    \caption{Mean $\Delta PA$ values (from a full period) for different models in our library, overlaid with our empirical fitting relation. The bottom panel shows residuals (percentage difference between $\Delta PA$ and $\delta PA_{0\rightarrow1}$), where for every radius the spin values have been expanded horizontally for clarity.}
    \label{fig:_fitK75}
\end{figure*}

Figure \ref{fig:_fitK75} illustrates the performance of this empirical relation. It compares mean $\Delta PA$ values (to avoid $i$ dependence) obtained from simulations for a certain set of parameters with those predicted by Equation \ref{eq:empirical}. The plots have been created to show models with fixed $\kc$ ($\kc = 0.75$ in this case) while varying ($r_{\textrm{hs}}$ and $a*$). The bottom panel shows the flat difference between the two $\Delta PA$ values, which remains below $5^\circ$ in all but two cases ($r_{\textrm{hs}}=4,\,a_*=-0.5,\,-0.9$) where it goes up to $10^\circ$. 

In the following section, we demonstrate its application through explicit examples.

\subsection{Mock {$\Delta PA$} {observation}}

Equation~\ref{eq:empirical} can be solved analytically for $a_*$, but the resulting expression is cumbersome and impractical to present here. Instead, a more convenient approach is to substitute observed values and solve for $a_*$. In our case, we employ a Bayesian solver using \dynesty, a nested sampling tool introduced by \cite{Speagle:2019ivv} and further developed by \cite{sergey_koposov_2023_7995596}.

This choice for a Bayesian approach is motivated by two key advantages: 1) it provides a natural framework for incorporating observational uncertainties through Gaussian priors with specified standard deviations, and 2) it yields a posterior distribution for the spin parameter, inherently accounting for measurement uncertainties. By assuming a uniform prior on spin, $a_* \in (-1,1)$, the resulting posterior directly reflects the inferred constraints on $a_*$.
For the likelihood calculation, we used: 

\begin{equation}
\mathcal{L}(\vec{p})=-\frac{\left[\Delta PA-\hat{\Delta PA}(\vec{p})\right]^2}{2\sigma_{\Delta PA}^2}\,,
    \label{eq:lklhd}
\end{equation}
where $\vec{p}$ is the vector of the model parameters ($ P,\, r_{\textrm{hs}},\,a_*$), $\Delta PA$ is the observed value, while $\hat{\Delta PA}$ is the predicted value from our empirical relation in Equation \ref{eq:empirical}. The uncertainty $\sigma_{\Delta PA}$ is taken from observations. 

Note that $P$ and $\rs$ are observed with their own uncertainties ($\sigma_P,\,\sigma_{\rs}$) but still re-sampled using their Gaussian distributions as prior (including the full parameter space). This is a similar approach to including all three observables in the $\mathcal{L}$ function and only sampling $a_*$, but better, because it extracts more information from the system, since the posteriors can have shifted peaks from the mean of the observation. It also provides a more intuitive framework, since it samples 3 values ($\rs,\,P$ and $a_*$) and extracts $\Delta PA$ that is then compared to the observation.  

Given the simplicity of the task, we employ dynamic nested sampling with 1000 live points, one additional batch, and default walker and bound settings. The computation takes approximately one minute on a standard {\tt Jupyter} notebook running on a typical laptop. The script for this analysis is publicly available at \href{https://github.com/ArYfantis/dyfit_dpa}{Github} \footnote{\url{https://github.com/ArYfantis/dyfit_dpa}}.

We simulate observations in which two bright spots can be identified on the screen, corresponding to the primary and secondary images of a hot spot. In terms of accuracy, the ring diameter was measured as $\hat{d} = 51.9 \pm 2 \,\mu$as in \cite{eht:2022_paperIV}. Based on this, we assume a comparable level of accuracy for $r_{\textrm{hs}}$ ($\pm 0.5,M$) and $\Delta PA$ ($\pm 5^\circ$). Using these uncertainties, we construct three models, summarized in Table \ref{tab:mock}. The errors of $\Delta PA$ are used for the likelihood calculation of Equation \ref{eq:lklhd}, while the errors of the other two parameters are used to define the prior space as a Gaussian distributions. 

\begin{table*}
  \caption{Mock observation parameters, including $a_*$ from our fits.}
    \centering
    \begin{tabular}{c|ccc | ccc}
         \toprule
         \multicolumn{4}{c|}{Observation Parameters} & \multicolumn{3}{c}{Spin Estimation} \vspace{0.2em}\\
         \midrule
         test & $\Delta PA[^\circ]$ & $r$[M] & $P$[min] & $a_*$ & $ 2\sigma_{a_*}$ & truth \vspace{0.2em}\\
         \midrule
         1 &  $140 \pm 5$ & $9 \pm 0.5$ & $90 \pm 3$ & $-0.16$ & $[-0.48,0.16]$ & $-0.18$\\
         2 & $172 \pm 5$ & $5.5 \pm 0.5$ & $54 \pm 3$ & $0.90$ & $[0.67,1]$ & $0.95$ \\
         3 &  $80 \pm 5$ & $4.5 \pm 0.5$ & $30 \pm 3$ & $-0.71$ & $[-0.97,-0.36]$ & $-0.78$\\
        \bottomrule
    \end{tabular}
     \tablefoot{The errors on the observational values are reported as $1\sigma$ Gaussian standard deviations. Note that $r$ and $P$ are parameters of the hot spot model, but are also directly constrained in our synthetic observations, so we resample them during fitting with a Gaussian prior given by the measurement.}
      \label{tab:mock}
\end{table*}

Before presenting the results, there is a subtle point from our empirical relation (Equation \ref{eq:empirical}) that is worth clarifying. The way we model orbital velocity  facilitates comparisons with numerical and analytical disk solutions; however, $\kc$ itself is not directly observable. Instead, observations provide measurements of $P$ and $r_{\textrm{hs}}$, which can be related to $\kc$ and $a_*$ through Equation \ref{eq:P}. This implies that we effectively have two equations with two unknowns. 

\begin{figure*}
    \centering
    \includegraphics[width=0.471\linewidth,trim={0.2cm 0.0cm 0 0.2cm},clip]{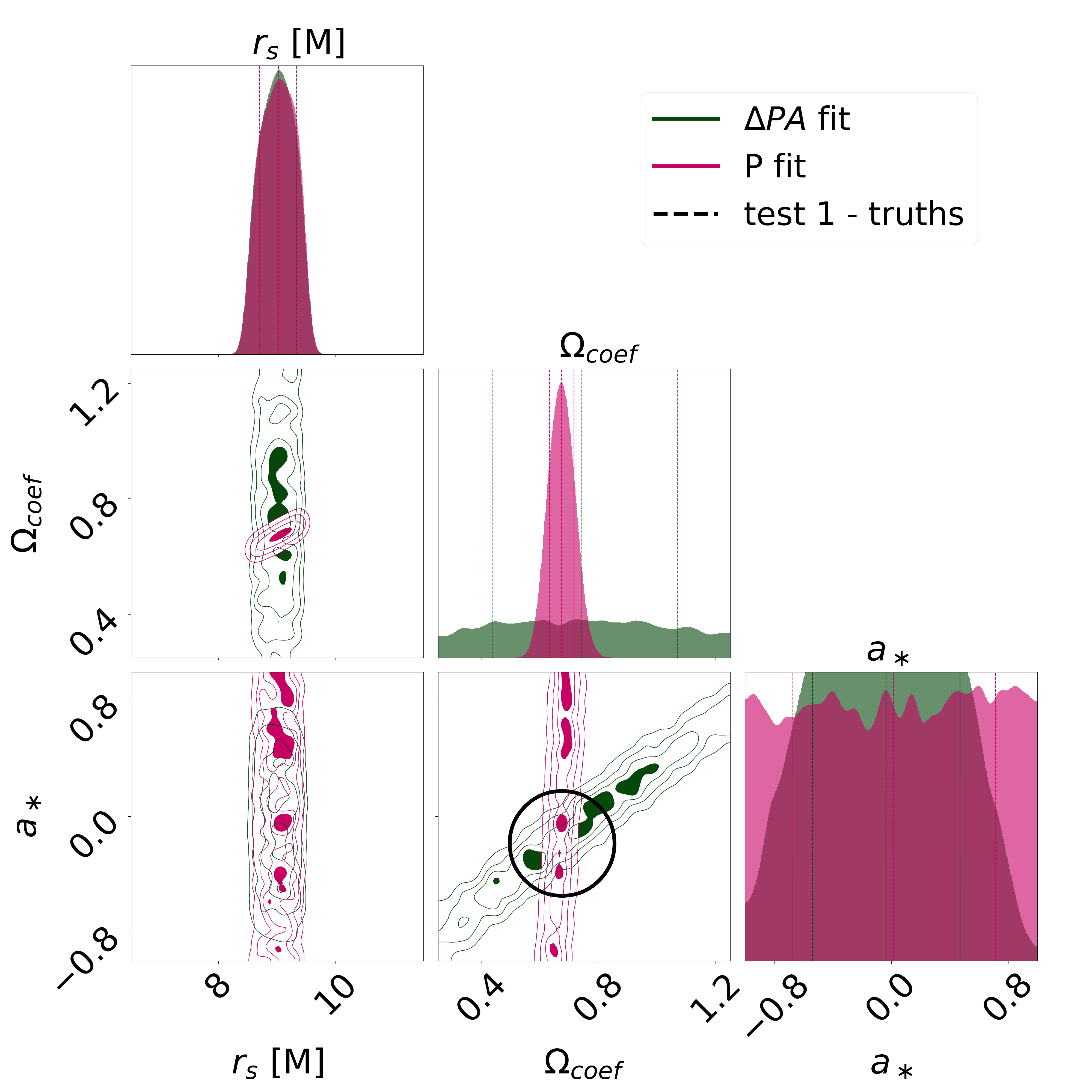}
    \includegraphics[width=0.471\linewidth,trim={0.2cm 0.0cm 0 0.2cm},clip]{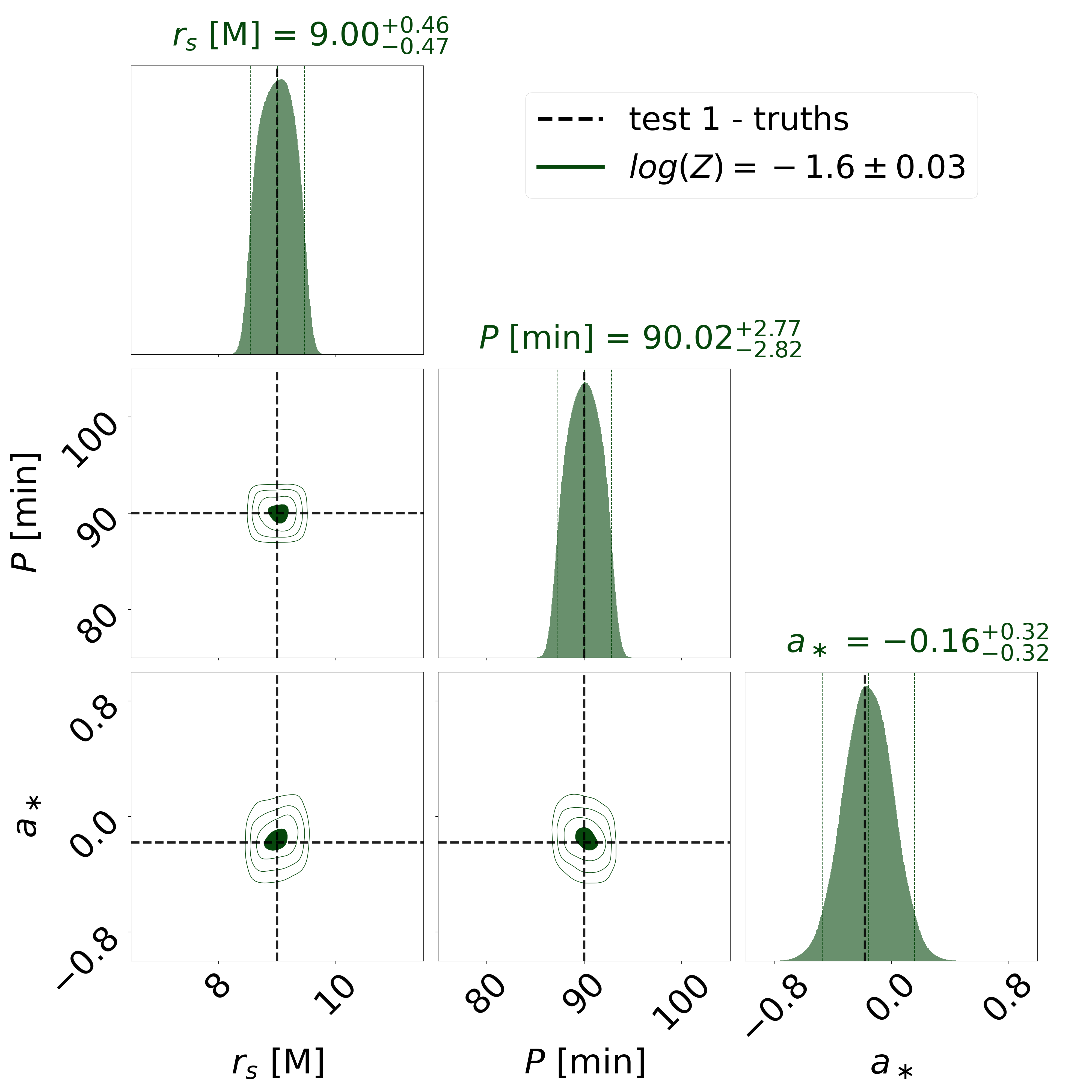} 

     \caption{A Bayesian fit using the mock observations with uncertainties from Table \ref{tab:mock}, and the empirical relation of Equation \ref{eq:empirical}. Left side shows the fits using Equation \ref{eq:empirical} and \ref{eq:P} (Period) separately, while right side shows the the fit when combining both.}
    \label{fig:mock_joint}
\end{figure*}

To illustrate this, we present Figure~\ref{fig:mock_joint}, where we use test 1 as a template of observation and perform two different estimations. In the left case, we estimate $\kc$ and $a_*$ using two different fits: the ``$\Delta PA$ fit,'' which relies on our empirical relation (Equation \ref{eq:empirical}), and the ``$P$ fit,'' which is based on Equation \ref{eq:P}. The ``$P$ fit'' constrains $\kc$ but leaves $a_*$ completely unconstrained, whereas the ``$\Delta PA$ fit'' has no direct constraint on $\kc$ and only weakly excludes the most extreme spin values.

The bottom-middle panel on the left of Figure~\ref{fig:mock_joint} shows the two-dimensional posterior distribution of $\kc$ and $a_*$. Although each individual fit allows $a_*$ to span nearly the full prior range, their overlap (highlighted by a black circle) demonstrates how combining both distributions leads to a well-defined solution. This provides a clear validation of our method’s effectiveness.

Instead of using a two-step approach that involves overlapping fits of the two equations, we can incorporate the exploration of $\kc$ within a single algorithm, as shown on the right side of Figure~\ref{fig:mock_joint}. In this case, $\kc$ is directly obtained from Equation~\ref{eq:Kc} and substituted into the empirical relation. Consequently, it is not treated as a free parameter in the prior and does not appear in the posteriors. Naturally, given the values of $r_{\textrm{hs}}$, $P$, and $a_*$, $\kc$ is fully determined. As expected from the previous exercise the spin parameter is nicely constrained at $a_*=-0.16 \pm 0.32$ with $2\sigma$ confidence.

\begin{figure*}
    \centering
     \includegraphics[width=0.471\linewidth,trim={0.2cm 0.0cm 0 0.2cm},clip]{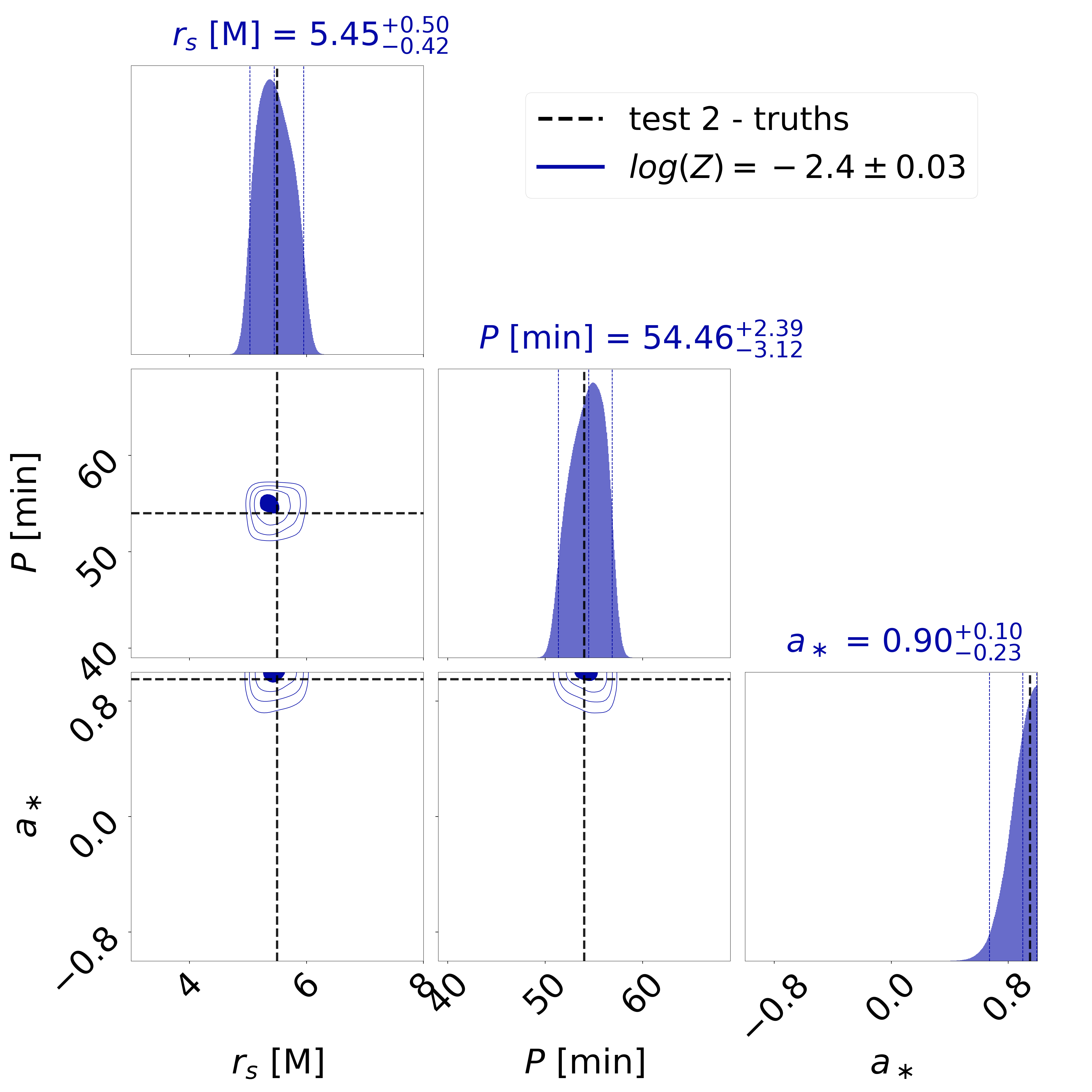} 
    \includegraphics[width=0.471\linewidth,trim={0.2cm 0.0cm 0 0.2cm},clip]{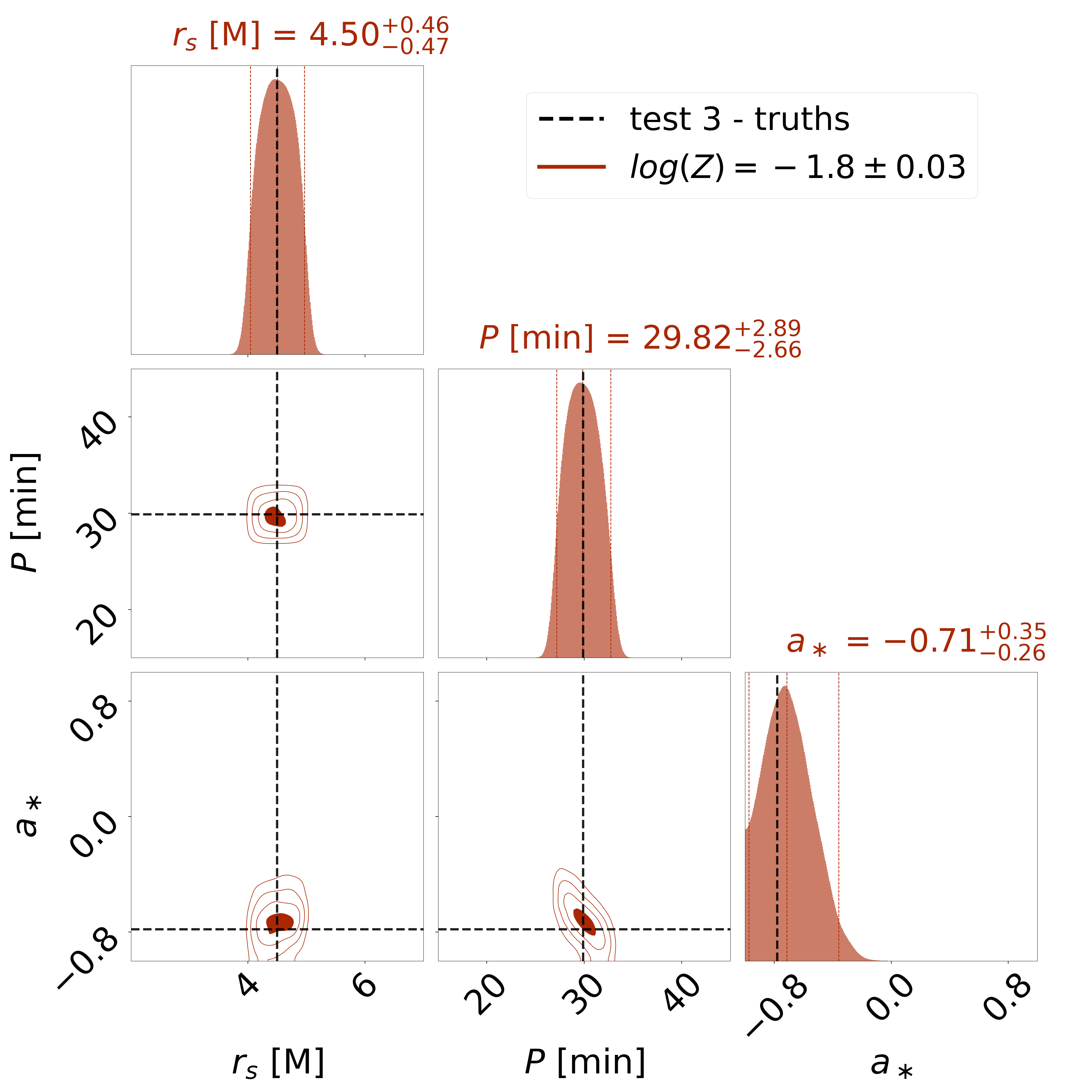} 
    
    \caption{A Bayesian fit using the mock observations with uncertainties from Table \ref{tab:mock}, and the empirical relation of Equation \ref{eq:empirical}. Note, that only $a_*$ is the true constrain here, while $\rs$ and $P$ are re-sampled for error propagation.}
    \label{fig:mock23}
\end{figure*}

The results from tests ``2" and ``3" are presented in Figure \ref{fig:mock23}. The overall behavior remains consistent with test 1, with one notable difference: the constraining power improves when the spin values are closer to the boundaries. This is evident in the case of $a_* = 0.94^{+0.06}_{-0.18}$, where the uncertainties are less than half those of tests 1 and 3. This behavior is expected, as the true value lies near the boundaries, but it is worth highlighting.

\section{Discussion and conclusions}\label{sec:discussion}
The measurement of a black hole's spin has occupied astrophysicists for decades, even before the first black hole image from the EHTC \citep{eht:2019_paperI}. With upcoming upgrades to the EHTC and ngEHT facilities, advancements in imaging algorithms, and the development of new projects such as BHEX, novel opportunities for spin estimation have emerged (and will continue to do so). This work explores one such avenue, leveraging the interpretation of flaring events around supermassive black holes, particularly Sgr A*, as emission from a well-localized region, a hot spot. This emission is expected to be sufficiently dominant, thus making the secondary image of the hot spot detectable, opening new possibilities for probing the near-horizon spacetime. Among the potential observables, we identify the difference in position angle between the primary ($n=0$) and secondary ($n=1$) hot spot images ($\Delta\textrm{PA}$) as a key diagnostic tool.

In this work, we present three key results that facilitate the use of this observable:

\begin{enumerate}
    \item We derive the correction to the complete elliptical integrals in the analytic relation for $\delta \textrm{PA}$, enabling the calculation of $\delta \textrm{PA}_{0\rightarrow1}$ in sub-degree accuracy, instead of the commonly used $n\rightarrow\infty$ limit. Additionally, we propose a simplified approximation to the analytic solution that deviates by at most $5^\circ$ at $a_* \rightarrow 1$ for the same $\delta \textrm{PA}_{0\rightarrow1}$ quantity.   

    \item We highlight a distinct effect of orbiting hot spots concerning the system’s inclination angle ($i$). Specifically, when tracking the values of $\Delta \textrm{PA}$ over a full orbit of a moving hot spot, we find that the mean value of $\Delta \textrm{PA}$ remains unchanged (sub-degree for all cases apart for $\rs=4,\,a_*=-0.9$ where deviation is three degrees) for inclinations up to $i \sim 30^\circ$, with  deviation of $\sim5^\circ$ emerging at $i \sim 60^\circ$. This finding  significantly simplifies our method, as otherwise, even at $i \sim 25^\circ$ --the expected inclination for \sgra-- the fluctuations of $\Delta \textrm{PA}$ would require a full curve analysis, and thus a more complicated comparison with models. 

    \item After studying a large library of more than 900 models generated using the GRRT code \ipole, we achieved our primary goal: constructing a simple empirical relation for estimating spin based on $\Delta \textrm{PA}$ observations. This method matches the analytic solution to within $5^\circ$ in most cases, with a maximum deviation of $10^\circ$ for $r_{\textrm{hs}} \leq 4$ M. To further demonstrate the viability of this method, we conducted three tests, using mock observations with realistic uncertainties, and successfully extracted the spin in all cases, with a maximum uncertainty of $\pm 0.3$ for $2\sigma$ confidence.

\end{enumerate}

In the case of an observation that does not capture a full period but for instance $50\%$ of it, we can still yield valuable constraints with a more extensive analysis. One approach is to assume an inclination angle ($i\sim 25^\circ$), and compare the $\Delta PA$ curve for several candidate values of the position angle ($PA$) of the BH spin axis on the sky. This is necessary in order to model the correct phase of the curve. Independent estimates of $PA$ include $PA\sim130^\circ$ from \cite{ball:2021} and \cite{yfantis24b}, while \cite{W22,G23,yfantis24a} are consistent with $PA\sim180^\circ$. From that we could extract the mean $\Delta PA$ and continue our moethod. A more robust method would be to create an adaptive Bayesian algorithm utilizing \bipole, similarly to \cite{yfantis24b,yfantis24a}, and explore the full parameter space, jointly fitting for $i,\,PA,\,\rs,\,\kc$ and $a_*$. While this approach requires significantly more computational time (up to ten days, compared to one minute for the current method), it yields a comprehensive posterior over all relevant parameters.

Looking forward, further refinement of this method could focus on expanding the parameter space to include in-falling or out-going orbits; this expansion would more readily reproduce equatorial motions seen in GRMHD. Another interesting addition would be the inclusion of non-equatorial orbits; these could address a big question in the tracking of flaring structures, regarding the mechanism for hot-spot creation, with off-equatorial spots often related to plamsoids rather than flux tubes. Another interesting avenue would be the creation of a machine learning algorithm using an expanded library that treats hot spot observational data more generally, including potentially partial sampling of the time-dependent astrometry for both images ($n=0,1$) and extract all the information for the system ($i$, $r_{\textrm{hs}}$, $\kc$, radial velocities, etc.).

\begin{acknowledgements}
We thank the anonymous EHT internal reviewer for their comments. This publication is a part of the project Dutch Black Hole Consortium (with project number NWA 1292.19.202) of the research program of the National Science Agenda which is financed by the Dutch Research Council (NWO).
\end{acknowledgements}

\balance
\bibliographystyle{aa} 
\bibliography{library} 

\begin{thebibliography}{58}
\expandafter\ifx\csname natexlab\endcsname\relax\def\natexlab#1{#1}\fi

\bibitem[{{Aimar} {et~al.}(2023){Aimar}, {Dmytriiev}, {Vincent}, {El Mellah}, {Paumard}, {Perrin}, \& {Zech}}]{aimar_2023_plasmoid}
{Aimar}, N., {Dmytriiev}, A., {Vincent}, F.~H., {et~al.} 2023, \aap, 672, A62

\bibitem[{{Antonopoulou} \& {Nathanail}(2024)}]{Antonopoulou24}
{Antonopoulou}, E. \& {Nathanail}, A. 2024, \aap, 690, A240

\bibitem[{{Ball} {et~al.}(2021){Ball}, {{\"O}zel}, {Christian}, {Chan}, \& {Psaltis}}]{ball:2021}
{Ball}, D., {{\"O}zel}, F., {Christian}, P., {Chan}, C.-K., \& {Psaltis}, D. 2021, \apj, 917, 8

\bibitem[{{Bardeen} {et~al.}(1972){Bardeen}, {Press}, \& {Teukolsky}}]{Bardeen_BH_1972}
{Bardeen}, J.~M., {Press}, W.~H., \& {Teukolsky}, S.~A. 1972, ApJ, 178, 347

\bibitem[{{Beckwith} \& {Done}(2005)}]{Beckwith_2005}
{Beckwith}, K. \& {Done}, C. 2005, \mnras, 359, 1217

\bibitem[{{Broderick} \& {Loeb}(2006)}]{Brod-Loeb2006}
{Broderick}, A.~E. \& {Loeb}, A. 2006, \mnras, 367, 905

\bibitem[{{Claudel} {et~al.}(2001){Claudel}, {Virbhadra}, \& {Ellis}}]{Claude2001}
{Claudel}, C.-M., {Virbhadra}, K.~S., \& {Ellis}, G.~F.~R. 2001, Journal of Mathematical Physics, 42, 818

\bibitem[{{Conroy} {et~al.}(2023){Conroy}, {Baub{\"o}ck}, {Dhruv}, {Lee}, {Broderick}, {Chan}, {Georgiev}, {Joshi}, {Prather}, \& {Gammie}}]{Conroy:2023kec}
{Conroy}, N.~S., {Baub{\"o}ck}, M., {Dhruv}, V., {et~al.} 2023, \apj, 951, 46

\bibitem[{{Dexter} {et~al.}(2020){Dexter}, {Tchekhovskoy}, {Jim{\'e}nez-Rosales}, {Ressler}, {Baub{\"o}ck}, {Dallilar}, {de Zeeuw}, {Eisenhauer}, {von Fellenberg}, {Gao}, {Genzel}, {Gillessen}, {Habibi}, {Ott}, {Stadler}, {Straub}, \& {Widmann}}]{dexter20}
{Dexter}, J., {Tchekhovskoy}, A., {Jim{\'e}nez-Rosales}, A., {et~al.} 2020, \mnras, 497, 4999

\bibitem[{{Doeleman} {et~al.}(2023){Doeleman}, {Barrett}, {Blackburn}, {Bouman}, {Broderick}, {Chaves}, {Fish}, {Fitzpatrick}, {Freeman}, {Fuentes}, {G{\'o}mez}, {Haworth}, {Houston}, {Issaoun}, {Johnson}, {Kettenis}, {Loinard}, {Nagar}, {Narayanan}, {Oppenheimer}, {Palumbo}, {Patel}, {Pesce}, {Raymond}, {Roelofs}, {Srinivasan}, {Tiede}, {Weintroub}, \& {Wielgus}}]{Doeleman_2023}
{Doeleman}, S.~S., {Barrett}, J., {Blackburn}, L., {et~al.} 2023, Galaxies, 11, 107

\bibitem[{{EHTC} {et~al.}(2024{\natexlab{a}}){EHTC}, {Akiyama}, {Alberdi}, {Alef}, {Algaba}, {Anantua}, {Asada}, {Azulay}, {Bach}, {Baczko}, {Ball}, {Balokovi{\'c}}, {Bandyopadhyay}, {Barrett}, {Baub{\"o}ck}, {Benson}, {Bintley}, {Blackburn}, {Blundell}, {Bouman}, {Bower}, {Boyce}, {Bremer}, {Brinkerink}, {Brissenden}, {Britzen}, {Broderick}, {Broguiere}, {Bronzwaer}, {Bustamante}, {Byun}, {Carlstrom}, {Ceccobello}, {Chael}, {Chan}, {Chang}, {Chatterjee}, {Chatterjee}, {Chen}, {Chen}, {Cheng}, {Cho}, {Christian}, {Conroy}, {Conway}, {Cordes}, {Crawford}, {Crew}, {Cruz-Osorio}, {Cui}, {Dahale}, {Davelaar}, {De Laurentis}, {Deane}, {Dempsey}, {Desvignes}, {Dexter}, {Dhruv}, {Dihingia}, {Doeleman}, {Dougall}, {Dzib}, {Eatough}, {Emami}, {Falcke}, {Farah}, {Fish}, {Fomalont}, {Ford}, {Foschi}, {Fraga-Encinas}, {Freeman}, {Friberg}, {Fromm}, {Fuentes}, {Galison}, {Gammie}, {Garc{\'\i}a}, {Gentaz}, {Georgiev}, {Goddi}, {Gold}, {G{\'o}mez-Ruiz}, {G{\'o}mez}, {Gu}, {Gurwell}, {Hada}, {Haggard}, {Haworth}, {Hecht},
  {Hesper}, {Heumann}, {Ho}, {Ho}, {Honma}, {Huang}, {Huang}, {Hughes}, {Ikeda}, {Impellizzeri}, {Inoue}, {Issaoun}, {James}, {Jannuzi}, {Janssen}, {Jeter}, {Jiang}, {Jim{\'e}nez-Rosales}, {Johnson}, {Jorstad}, {Joshi}, {Jung}, {Karami}, {Karuppusamy}, {Kawashima}, {Keating}, {Kettenis}, {Kim}, {Kim}, {Kim}, {Kim}, {Kino}, {Koay}, {Kocherlakota}, {Kofuji}, {Koch}, {Koyama}, {Kramer}, {Kramer}, {Kramer}, {Krichbaum}, {Kuo}, {La Bella}, {Lauer}, {Lee}, {Lee}, {Leung}, {Levis}, {Li}, {Lico}, {Lindahl}, {Lindqvist}, {Lisakov}, {Liu}, {Liu}, {Liuzzo}, {Lo}, {Lobanov}, {Loinard}, {Lonsdale}, {Lowitz}, {Lu}, {MacDonald}, {Mao}, {Marchili}, {Markoff}, {Marrone}, {Marscher}, {Mart{\'\i}-Vidal}, {Matsushita}, {Matthews}, {Medeiros}, {Menten}, {Michalik}, {Mizuno}, {Mizuno}, {Moran}, {Moriyama}, {Moscibrodzka}, {Mulaudzi}, {M{\"u}ller}, {M{\"u}ller}, {Mus}, {Musoke}, {Myserlis}, {Nadolski}, {Nagai}, {Nagar}, {Nakamura}, {Narayanan}, {Natarajan}, {Nathanail}, {Fuentes}, {Neilsen}, {Neri}, {Ni}, {Noutsos}, {Nowak}, {Oh},
  {Okino}, {Olivares}, {Ortiz-Le{\'o}n}, {Oyama}, {{\"O}zel}, {Palumbo}, {Paraschos}, {Park}, {Parsons}, {Patel}, \& {Pen}}]{eht:2024b}
{EHTC}, {Akiyama}, K., {Alberdi}, A., {et~al.} 2024{\natexlab{a}}, \apjl, 964, L26

\bibitem[{{EHTC} {et~al.}(2024{\natexlab{b}}){EHTC}, {Akiyama}, {Alberdi}, {Alef}, {Algaba}, {Anantua}, {Asada}, {Azulay}, {Bach}, {Baczko}, {Ball}, {Balokovi{\'c}}, {Bandyopadhyay}, {Barrett}, {Baub{\"o}ck}, {Benson}, {Bintley}, {Blackburn}, {Blundell}, {Bouman}, {Bower}, {Boyce}, {Bremer}, {Brissenden}, {Britzen}, {Broderick}, {Broguiere}, {Bronzwaer}, {Bustamante}, {Carlstrom}, {Chael}, {Chan}, {Chang}, {Chatterjee}, {Chatterjee}, {Chen}, {Chen}, {Cheng}, {Cho}, {Christian}, {Conroy}, {Conway}, {Crawford}, {Crew}, {Cruz-Osorio}, {Cui}, {Dahale}, {Davelaar}, {De Laurentis}, {Deane}, {Dempsey}, {Desvignes}, {Dexter}, {Dhruv}, {Dihingia}, {Doeleman}, {Dzib}, {Eatough}, {Emami}, {Falcke}, {Farah}, {Fish}, {Fomalont}, {Ford}, {Foschi}, {Fraga-Encinas}, {Freeman}, {Friberg}, {Fromm}, {Fuentes}, {Galison}, {Gammie}, {Garc{\'\i}a}, {Gentaz}, {Georgiev}, {Goddi}, {Gold}, {G{\'o}mez-Ruiz}, {G{\'o}mez}, {Gu}, {Gurwell}, {Hada}, {Haggard}, {Hesper}, {Heumann}, {Ho}, {Ho}, {Honma}, {Huang}, {Huang}, {Hughes}, {Ikeda},
  {Violette Impellizzeri}, {Inoue}, {Issaoun}, {James}, {Jannuzi}, {Janssen}, {Jeter}, {Jiang}, {Jim{\'e}nez-Rosales}, {Johnson}, {Jorstad}, {Jones}, {Joshi}, {Jung}, {Karuppusamy}, {Kawashima}, {Keating}, {Kettenis}, {Kim}, {Kim}, {Kim}, {Kim}, {Kino}, {Koay}, {Kocherlakota}, {Kofuji}, {Koch}, {Koyama}, {Kramer}, {Kramer}, {Kramer}, {Krichbaum}, {Kuo}, {La Bella}, {Lee}, {Levis}, {Li}, {Lico}, {Lindahl}, {Lindqvist}, {Lisakov}, {Liu}, {Liu}, {Liuzzo}, {Lo}, {Lobanov}, {Loinard}, {Lonsdale}, {Lowitz}, {Lu}, {MacDonald}, {Mao}, {Marchili}, {Markoff}, {Marrone}, {Marscher}, {Mart{\'\i}-Vidal}, {Matsushita}, {Matthews}, {Medeiros}, {Menten}, {Mizuno}, {Mizuno}, {Montgomery}, {Moran}, {Moriyama}, {Moscibrodzka}, {Mulaudzi}, {M{\"u}ller}, {M{\"u}ller}, {Mus}, {Musoke}, {Myserlis}, {Nagai}, {Nagar}, {Nakamura}, {Narayanan}, {Natarajan}, {Nathanail}, {Fuentes}, {Neilsen}, {Ni}, {Nowak}, {Oh}, {Okino}, {Olivares}, {Oyama}, {{\"O}zel}, {Palumbo}, {Paraschos}, {Park}, {Parsons}, {Patel}, {Pen}, {Pesce}, {Pi{\'e}tu},
  {PopStefanija}, {Porth}, {Prather}, {Psaltis}, {Pu}, {Ramakrishnan}, {Rao}, {Rawlings}, {Raymond}, {Rezzolla}, {Ricarte}, \& {Ripperda}}]{eht_2024a}
{EHTC}, {Akiyama}, K., {Alberdi}, A., {et~al.} 2024{\natexlab{b}}, \aap, 681, A79

\bibitem[{{EHTC} {et~al.}(2022{\natexlab{a}}){EHTC}, {Akiyama}, {Alberdi}, {Alef}, {Algaba}, {Anantua}, {Asada}, {Azulay}, {Bach}, {Baczko}, {Ball}, {Balokovi{\'c}}, {Barrett}, {Baub{\"o}ck}, {Benson}, {Bintley}, {Blackburn}, {Blundell}, {Bouman}, {Bower}, {Boyce}, {Bremer}, {Brinkerink}, {Brissenden}, {Britzen}, {Broderick}, {Broguiere}, {Bronzwaer}, {Bustamante}, {Byun}, {Carlstrom}, {Ceccobello}, {Chael}, {Chan}, {Chatterjee}, {Chatterjee}, {Chen}, {Chen}, {Cheng}, {Cho}, {Christian}, {Conroy}, {Conway}, {Cordes}, {Crawford}, {Crew}, {Cruz-Osorio}, {Cui}, {Davelaar}, {De Laurentis}, {Deane}, {Dempsey}, {Desvignes}, {Dexter}, {Dhruv}, {Doeleman}, {Dougal}, {Dzib}, {Eatough}, {Emami}, {Falcke}, {Farah}, {Fish}, {Fomalont}, {Ford}, {Fraga-Encinas}, {Freeman}, {Friberg}, {Fromm}, {Fuentes}, {Galison}, {Gammie}, {Garc{\'\i}a}, {Gentaz}, {Georgiev}, {Goddi}, {Gold}, {G{\'o}mez-Ruiz}, {G{\'o}mez}, {Gu}, {Gurwell}, {Hada}, {Haggard}, {Haworth}, {Hecht}, {Hesper}, {Heumann}, {Ho}, {Ho}, {Honma}, {Huang}, {Huang},
  {Hughes}, {Ikeda}, {Impellizzeri}, {Inoue}, {Issaoun}, {James}, {Jannuzi}, {Janssen}, {Jeter}, {Jiang}, {Jim{\'e}nez-Rosales}, {Johnson}, {Jorstad}, {Joshi}, {Jung}, {Karami}, {Karuppusamy}, {Kawashima}, {Keating}, {Kettenis}, {Kim}, {Kim}, {Kim}, {Kim}, {Kino}, {Koay}, {Kocherlakota}, {Kofuji}, {Koch}, {Koyama}, {Kramer}, {Kramer}, {Krichbaum}, {Kuo}, {La Bella}, {Lauer}, {Lee}, {Lee}, {Leung}, {Levis}, {Li}, {Lico}, {Lindahl}, {Lindqvist}, {Lisakov}, {Liu}, {Liu}, {Liuzzo}, {Lo}, {Lobanov}, {Loinard}, {Lonsdale}, {Lu}, {Mao}, {Marchili}, {Markoff}, {Marrone}, {Marscher}, {Mart{\'\i}-Vidal}, {Matsushita}, {Matthews}, {Medeiros}, {Menten}, {Michalik}, {Mizuno}, {Mizuno}, {Moran}, {Moriyama}, {Moscibrodzka}, {M{\"u}ller}, {Mus}, {Musoke}, {Myserlis}, {Nadolski}, {Nagai}, {Nagar}, {Nakamura}, {Narayan}, {Narayanan}, {Natarajan}, {Nathanail}, {Fuentes}, {Neilsen}, {Neri}, {Ni}, {Noutsos}, {Nowak}, {Oh}, {Okino}, {Olivares}, {Ortiz-Le{\'o}n}, {Oyama}, {Palumbo}, {Paraschos}, {Park}, {Parsons}, {Patel}, {Pen},
  {Pesce}, {Pi{\'e}tu}, {Plambeck}, {PopStefanija}, {Porth}, {P{\"o}tzl}, {Prather}, {Preciado-L{\'o}pez}, {Pu}, {Ramakrishnan}, {Rao}, {Rawlings}, {Raymond}, {Rezzolla}, {Ricarte}, {Ripperda}, {Roelofs}, {Rogers}, {Ros}, {Romero-Ca{\~n}izales}, {Roshanineshat}, {Rottmann}, {Roy}, {Ruiz}, {Ruszczyk}, {Rygl}, {S{\'a}nchez}, {S{\'a}nchez-Arg{\"u}elles}, {S{\'a}nchez-Portal}, {Sasada}, {Satapathy}, {Savolainen}, {Schloerb}, {Schonfeld}, {Schuster}, {Shao}, {Shen}, {Small}, {Sohn}, {SooHoo}, {Souccar}, {Sun}, {Tazaki}, {Tetarenko}, {Tiede}, {Tilanus}, {Titus}, {Torne}, {Traianou}, {Trent}, {Trippe}, {Turk}, {van Bemmel}, {van Langevelde}, {van Rossum}, {Vos}, {Wagner}, {Ward-Thompson}, {Wardle}, {Weintroub}, {Wex}, {Wharton}, {Wielgus}, {Wiik}, {Witzel}, {Wondrak}, {Wong}, {Wu}, {Yamaguchi}, {Yoon}, {Young}, {Young}, {Younsi}, {Yuan}, {Yuan}, {Zensus}, {Zhang}, {Zhao}, {Zhao}, \& {Chang}}]{eht:2022_paperIV}
{EHTC}, {Akiyama}, K., {Alberdi}, A., {et~al.} 2022{\natexlab{a}}, \apjl, 930, L15

\bibitem[{{EHTC} {et~al.}(2019{\natexlab{a}}){EHTC}, {Akiyama}, {Alberdi}, {Alef}, {Asada}, {Azulay}, {Baczko}, {Ball}, {Balokovi{\'c}}, {Barrett}, {Bintley}, {Blackburn}, {Boland}, {Bouman}, {Bower}, {Bremer}, {Brinkerink}, {Brissenden}, {Britzen}, {Broderick}, {Broguiere}, {Bronzwaer}, {Byun}, {Carlstrom}, {Chael}, {Chan}, {Chatterjee}, {Chatterjee}, {Chen}, {Chen}, {Cho}, {Christian}, {Conway}, {Cordes}, {Crew}, {Cui}, {Davelaar}, {De Laurentis}, {Deane}, {Dempsey}, {Desvignes}, {Dexter}, {Doeleman}, {Eatough}, {Falcke}, {Fish}, {Fomalont}, {Fraga-Encinas}, {Friberg}, {Fromm}, {G{\'o}mez}, {Galison}, {Gammie}, {Garc{\'\i}a}, {Gentaz}, {Georgiev}, {Goddi}, {Gold}, {Gu}, {Gurwell}, {Hada}, {Hecht}, {Hesper}, {Ho}, {Ho}, {Honma}, {Huang}, {Huang}, {Hughes}, {Ikeda}, {Inoue}, {Issaoun}, {James}, {Jannuzi}, {Janssen}, {Jeter}, {Jiang}, {Johnson}, {Jorstad}, {Jung}, {Karami}, {Karuppusamy}, {Kawashima}, {Keating}, {Kettenis}, {Kim}, {Kim}, {Kim}, {Kino}, {Koay}, {Koch}, {Koyama}, {Kramer}, {Kramer}, {Krichbaum},
  {Kuo}, {Lauer}, {Lee}, {Li}, {Li}, {Lindqvist}, {Liu}, {Liuzzo}, {Lo}, {Lobanov}, {Loinard}, {Lonsdale}, {Lu}, {MacDonald}, {Mao}, {Markoff}, {Marrone}, {Marscher}, {Mart{\'\i}-Vidal}, {Matsushita}, {Matthews}, {Medeiros}, {Menten}, {Mizuno}, {Mizuno}, {Moran}, {Moriyama}, {Moscibrodzka}, {Mul{\ensuremath{\ddot{}}}ler}, {Nagai}, {Nagar}, {Nakamura}, {Narayan}, {Narayanan}, {Natarajan}, {Neri}, {Ni}, {Noutsos}, {Okino}, {Olivares}, {Oyama}, {{\"O}zel}, {Palumbo}, {Patel}, {Pen}, {Pesce}, {Pi{\'e}tu}, {Plambeck}, {PopStefanija}, {Porth}, {Prather}, {Preciado-L{\'o}pez}, {Psaltis}, {Pu}, {Ramakrishnan}, {Rao}, {Rawlings}, {Raymond}, {Rezzolla}, {Ripperda}, {Roelofs}, {Rogers}, {Ros}, {Rose}, {Roshanineshat}, {Rottmann}, {Roy}, {Ruszczyk}, {Ryan}, {Rygl}, {S{\'a}nchez}, {S{\'a}nchez-Arguelles}, {Sasada}, {Savolainen}, {Schloerb}, {Schuster}, {Shao}, {Shen}, {Small}, {Sohn}, {SooHoo}, {Tazaki}, {Tiede}, {Tilanus}, {Titus}, {Toma}, {Torne}, {Trent}, {Trippe}, {Tsuda}, {van Bemmel}, {van Langevelde}, {van Rossum},
  {Wagner}, {Wardle}, {Weintroub}, {Wex}, {Wharton}, {Wielgus}, {Wong}, {Wu}, {Young}, {Young}, {Younsi}, \& {Yuan}}]{eht:2019e}
{EHTC}, {Akiyama}, K., {Alberdi}, A., {et~al.} 2019{\natexlab{a}}, \apjl, 875, L5

\bibitem[{{EHTC} {et~al.}(2019{\natexlab{b}}){EHTC}, {Akiyama}, {Alberdi}, {Alef}, {Asada}, {Azulay}, {Baczko}, {Ball}, {Balokovi{\'c}}, {Barrett}, \& et~al.}]{eht:2019_paperI}
{EHTC}, {Akiyama}, K., {Alberdi}, A., {et~al.} 2019{\natexlab{b}}, \apjl, 875, L1

\bibitem[{{EHTC} {et~al.}(2022{\natexlab{b}}){EHTC}, {Alberdi}, {Alef}, {Algaba}, {Anantua}, {Asada}, {Azulay}, {Bach}, {Baczko}, {Ball}, {Balokovi{\'c}}, {Barrett}, {Baub{\"o}ck}, \& {Benson}}]{eht:2022_paperI}
{EHTC}, {Akiyama}, K., {Alberdi}, A., {Alef}, W., {et~al.} 2022{\natexlab{b}}, \apjl, 930, L12

\bibitem[{{EHTC} {et~al.}(2022{\natexlab{c}}){EHTC}, {Alberdi}, {Alef}, {Algaba}, {Anantua}, {Asada}, {Azulay}, {Bach}, {Baczko}, {Ball}, {Balokovi{\'c}}, {Barrett}, {Baub{\"o}ck}, \& {Benson}}]{eht:2022_paperII}
{EHTC}, {Akiyama}, K., {Alberdi}, A., {Alef}, W., {et~al.} 2022{\natexlab{c}}, \apjl, 930, L13

\bibitem[{{EHTC} {et~al.}(2022{\natexlab{d}}){EHTC}, {Alberdi}, {Alef}, {Algaba}, {Anantua}, {Asada}, {Azulay}, {Bach}, {Baczko}, {Ball}, {Balokovi{\'c}}, {Barrett}, {Baub{\"o}ck}, \& {Benson}}]{eht:2022_paperV}
{EHTC}, {Akiyama}, K., {Alberdi}, A., {Alef}, W., {et~al.} 2022{\natexlab{d}}, \apjl, 930, L16

\bibitem[{{Galison} {et~al.}(2024){Galison}, {Johnson}, {Lupsasca}, {Gravely}, \& {Berens}}]{BHEX2024b}
{Galison}, P., {Johnson}, M.~D., {Lupsasca}, A., {Gravely}, T., \& {Berens}, R. 2024, arXiv e-prints, arXiv:2406.11671

\bibitem[{{Gelles} {et~al.}(2021){Gelles}, {Himwich}, {Johnson}, \& {Palumbo}}]{Gelles:2021}
{Gelles}, Z., {Himwich}, E., {Johnson}, M.~D., \& {Palumbo}, D. C.~M. 2021, \prd, 104, 044060

\bibitem[{{Gralla} {et~al.}(2019){Gralla}, {Holz}, \& {Wald}}]{Gralla2019}
{Gralla}, S.~E., {Holz}, D.~E., \& {Wald}, R.~M. 2019, \prd, 100, 024018

\bibitem[{{Gralla} \& {Lupsasca}(2020{\natexlab{a}})}]{Gralla2020_lensing}
{Gralla}, S.~E. \& {Lupsasca}, A. 2020{\natexlab{a}}, \prd, 101, 044031

\bibitem[{{Gralla} \& {Lupsasca}(2020{\natexlab{b}})}]{Gralla2020}
{Gralla}, S.~E. \& {Lupsasca}, A. 2020{\natexlab{b}}, \prd, 101, 044031

\bibitem[{{Gralla} {et~al.}(2020){Gralla}, {Lupsasca}, \& {Marrone}}]{Gralla2020_grtest}
{Gralla}, S.~E., {Lupsasca}, A., \& {Marrone}, D.~P. 2020, \prd, 102, 124004

\bibitem[{{GRAVITY Collaboration} {et~al.}(2023){GRAVITY Collaboration}, {Abuter}, {Aimar}, {Amaro Seoane}, {Amorim}, {Baub{\"o}ck}, {Berger}, {Bonnet}, {Bourdarot}, {Brandner}, {Cardoso}, {Cl{\'e}net}, {Davies}, {de Zeeuw}, {Dexter}, {Drescher}, {Eckart}, {Eisenhauer}, {Feuchtgruber}, {Finger}, {F{\"o}rster Schreiber}, {Foschi}, {Garcia}, {Gao}, {Gelles}, {Gendron}, {Genzel}, {Gillessen}, {Hartl}, {Haubois}, {Haussmann}, {Hei{\ss}el}, {Henning}, {Hippler}, {Horrobin}, {Jochum}, {Jocou}, {Kaufer}, {Kervella}, {Lacour}, {Lapeyr{\`e}re}, {Le Bouquin}, {L{\'e}na}, {Lutz}, {Mang}, {More}, {Ott}, {Paumard}, {Perraut}, {Perrin}, {Pfuhl}, {Rabien}, {Ribeiro}, {Sadun Bordoni}, {Scheithauer}, {Shangguan}, {Shimizu}, {Stadler}, {Straub}, {Straubmeier}, {Sturm}, {Tacconi}, {Vincent}, {von Fellenberg}, {Widmann}, {Wielgus}, {Wieprecht}, {Wiezorrek}, \& {Woillez}}]{G23}
{GRAVITY Collaboration}, {Abuter}, R., {Aimar}, N., {et~al.} 2023, \aap, 677, L10

\bibitem[{{GRAVITY Collaboration} {et~al.}(2022){GRAVITY Collaboration}, {Abuter}, {Aimar}, {Amorim}, {Ball}, {Baub{\"o}ck}, {Berger}, {Bonnet}, {Bourdarot}, {Brandner}, {Cardoso}, {Cl{\'e}net}, {Dallilar}, {Davies}, {de Zeeuw}, {Dexter}, {Drescher}, {Eisenhauer}, {F{\"o}rster Schreiber}, {Foschi}, {Garcia}, {Gao}, {Gendron}, {Genzel}, {Gillessen}, {Habibi}, {Haubois}, {Hei{\ss}el}, {Henning}, {Hippler}, {Horrobin}, {Jochum}, {Jocou}, {Kaufer}, {Kervella}, {Lacour}, {Lapeyr{\`e}re}, {Le Bouquin}, {L{\'e}na}, {Lutz}, {Ott}, {Paumard}, {Perraut}, {Perrin}, {Pfuhl}, {Rabien}, {Shangguan}, {Shimizu}, {Scheithauer}, {Stadler}, {Stephens}, {Straub}, {Straubmeier}, {Sturm}, {Tacconi}, {Tristram}, {Vincent}, {von Fellenberg}, {Widmann}, {Wieprecht}, {Wiezorrek}, {Woillez}, {Yazici}, \& {Young}}]{Gravity2022}
{GRAVITY Collaboration}, {Abuter}, R., {Aimar}, N., {et~al.} 2022, \aap, 657, L12

\bibitem[{{GRAVITY Collaboration} {et~al.}(2020{\natexlab{a}}){GRAVITY Collaboration}, {Abuter}, {Amorim}, {Baub{\"o}ck}, {Berger}, {Bonnet}, {Brandner}, {Cardoso}, {Cl{\'e}net}, {de Zeeuw}, {Dallilar}, {Dexter}, {Eckart}, {Eisenhauer}, {F{\"o}rster Schreiber}, {Garcia}, {Gao}, {Gendron}, {Genzel}, {Gillessen}, {Habibi}, {Haubois}, {Henning}, {Hippler}, {Horrobin}, {Jim{\'e}nez-Rosales}, {Jochum}, {Jocou}, {Kaufer}, {Kervella}, {Lacour}, {Lapeyr{\`e}re}, {Le Bouquin}, {L{\'e}na}, {Nowak}, {Ott}, {Paumard}, {Perraut}, {Perrin}, {Pfuhl}, {Ponti}, {Rodriguez Coira}, {Shangguan}, {Scheithauer}, {Stadler}, {Straub}, {Straubmeier}, {Sturm}, {Tacconi}, {Vincent}, {von Fellenberg}, {Waisberg}, {Widmann}, {Wieprecht}, {Wiezorrek}, {Woillez}, {Yazici}, \& {Zins}}]{gravity:2020c}
{GRAVITY Collaboration}, {Abuter}, R., {Amorim}, A., {et~al.} 2020{\natexlab{a}}, \aap, 638, A2

\bibitem[{{GRAVITY Collaboration} {et~al.}(2018){GRAVITY Collaboration}, {Abuter}, {Amorim}, {Baub{\"o}ck}, {Berger}, {Bonnet}, {Brandner}, {Cl{\'e}net}, {Coud{\'e} Du Foresto}, {de Zeeuw}, {Deen}, {Dexter}, {Duvert}, {Eckart}, {Eisenhauer}, {F{\"o}rster Schreiber}, {Garcia}, {Gao}, {Gendron}, {Genzel}, {Gillessen}, {Guajardo}, {Habibi}, {Haubois}, {Henning}, {Hippler}, {Horrobin}, {Huber}, {Jim{\'e}nez-Rosales}, {Jocou}, {Kervella}, {Lacour}, {Lapeyr{\`e}re}, {Lazareff}, {Le Bouquin}, {L{\'e}na}, {Lippa}, {Ott}, {Panduro}, {Paumard}, {Perraut}, {Perrin}, {Pfuhl}, {Plewa}, {Rabien}, {Rodr{\'\i}guez-Coira}, {Rousset}, {Sternberg}, {Straub}, {Straubmeier}, {Sturm}, {Tacconi}, {Vincent}, {von Fellenberg}, {Waisberg}, {Widmann}, {Wieprecht}, {Wiezorrek}, {Woillez}, \& {Yazici}}]{gravity:2018}
{GRAVITY Collaboration}, {Abuter}, R., {Amorim}, A., {et~al.} 2018, \aap, 618, L10

\bibitem[{{GRAVITY Collaboration} {et~al.}(2020{\natexlab{b}}){GRAVITY Collaboration}, {Baub{\"o}ck}, {Dexter}, {Abuter}, {Amorim}, {Berger}, {Bonnet}, {Brandner}, {Cl{\'e}net}, {Coud{\'e} Du Foresto}, {de Zeeuw}, {Duvert}, {Eckart}, {Eisenhauer}, {F{\"o}rster Schreiber}, {Gao}, {Garcia}, {Gendron}, {Genzel}, {Gerhard}, {Gillessen}, {Habibi}, {Haubois}, {Henning}, {Hippler}, {Horrobin}, {Jim{\'e}nez-Rosales}, {Jocou}, {Kervella}, {Lacour}, {Lapeyr{\`e}re}, {Le Bouquin}, {L{\'e}na}, {Ott}, {Paumard}, {Perraut}, {Perrin}, {Pfuhl}, {Rabien}, {Rodriguez Coira}, {Rousset}, {Scheithauer}, {Stadler}, {Sternberg}, {Straub}, {Straubmeier}, {Sturm}, {Tacconi}, {Vincent}, {von Fellenberg}, {Waisberg}, {Widmann}, {Wieprecht}, {Wiezorrek}, {Woillez}, \& {Yazici}}]{gravityMichi}
{GRAVITY Collaboration}, {Baub{\"o}ck}, M., {Dexter}, J., {et~al.} 2020{\natexlab{b}}, \aap, 635, A143

\bibitem[{{GRAVITY Collaboration} {et~al.}(2020{\natexlab{c}}){GRAVITY Collaboration}, {Jim{\'e}nez-Rosales}, {Dexter}, {Widmann}, {Baub{\"o}ck}, {Abuter}, {Amorim}, {Berger}, {Bonnet}, {Brandner}, {Cl{\'e}net}, {de Zeeuw}, {Eckart}, {Eisenhauer}, {F{\"o}rster Schreiber}, {Garcia}, {Gao}, {Gendron}, {Genzel}, {Gillessen}, {Habibi}, {Haubois}, {Hei{\ss}el}, {Henning}, {Hippler}, {Horrobin}, {Jochum}, {Jocou}, {Kaufer}, {Kervella}, {Lacour}, {Lapeyr{\`e}re}, {Le Bouquin}, {L{\'e}na}, {Nowak}, {Ott}, {Paumard}, {Perraut}, {Perrin}, {Pfuhl}, {Rodr{\'\i}guez-Coira}, {Shangguan}, {Scheithauer}, {Stadler}, {Straub}, {Straubmeier}, {Sturm}, {Tacconi}, {Vincent}, {von Fellenberg}, {Waisberg}, {Wieprecht}, {Wiezorrek}, {Woillez}, {Yazici}, \& {Zins}}]{gravityAlejandra}
{GRAVITY Collaboration}, {Jim{\'e}nez-Rosales}, A., {Dexter}, J., {et~al.} 2020{\natexlab{c}}, \aap, 643, A56

\bibitem[{{Haggard} {et~al.}(2019){Haggard}, {Nynka}, {Mon}, {de la Cruz Hernandez}, {Nowak}, {Heinke}, {Neilsen}, {Dexter}, {Fragile}, {Baganoff}, {Bower}, {Corrales}, {Coti Zelati}, {Degenaar}, {Markoff}, {Morris}, {Ponti}, {Rea}, {Wilms}, \& {Yusef-Zadeh}}]{sgra_xflares2019}
{Haggard}, D., {Nynka}, M., {Mon}, B., {et~al.} 2019, \apj, 886, 96

\bibitem[{{Hamaus} {et~al.}(2009){Hamaus}, {Paumard}, {M{\"u}ller}, {Gillessen}, {Eisenhauer}, {Trippe}, \& {Genzel}}]{Hamaus2009}
{Hamaus}, N., {Paumard}, T., {M{\"u}ller}, T., {et~al.} 2009, \apj, 692, 902

\bibitem[{{Johannsen} \& {Psaltis}(2010)}]{JP2010}
{Johannsen}, T. \& {Psaltis}, D. 2010, \apj, 718, 446

\bibitem[{{Johnson} {et~al.}(2024){Johnson}, {Akiyama}, {Baturin}, {Bilyeu}, {Blackburn}, {Boroson}, {C{\'a}rdenas-Avenda{\~n}o}, {Chael}, {Chan}, {Chang}, {Cheimets}, {Chou}, {Doeleman}, {Farah}, {Galison}, {Gamble}, {Gammie}, {Gelles}, {G{\'o}mez}, {Gralla}, {Grimes}, {Gurvits}, {Hadar}, {Haworth}, {Hada}, {Hecht}, {Honma}, {Houston}, {Hudson}, {Issaoun}, {Jia}, {Jorstad}, {Kauffman}, {Kovalev}, {Kurczynski}, {Lafon}, {Lupsasca}, {Lehmensiek}, {Ma}, {Marrone}, {Marscher}, {Melnick}, {Narayan}, {Niinuma}, {Noble}, {Palmer}, {Palumbo}, {Paritsky}, {Peretz}, {Pesce}, {Plavin}, {Quataert}, {Rana}, {Ricarte}, {Roelofs}, {Shtyrkova}, {Sinclair}, {Small}, {Kumara}, {Srinivasan}, {Strominger}, {Tiede}, {Tong}, {Wang}, {Weintroub}, {Wielgus}, \& {Wong}}]{Johnson_2024}
{Johnson}, M.~D., {Akiyama}, K., {Baturin}, R., {et~al.} 2024, in Society of Photo-Optical Instrumentation Engineers (SPIE) Conference Series, Vol. 13092, Space Telescopes and Instrumentation 2024: Optical, Infrared, and Millimeter Wave, ed. L.~E. {Coyle}, S.~{Matsuura}, \& M.~D. {Perrin}, 130922D

\bibitem[{{Johnson} {et~al.}(2020){Johnson}, {Lupsasca}, {Strominger}, {Wong}, {Hadar}, {Kapec}, {Narayan}, {Chael}, {Gammie}, {Galison}, {Palumbo}, {Doeleman}, {Blackburn}, {Wielgus}, {Pesce}, {Farah}, \& {Moran}}]{Johnson2020}
{Johnson}, M.~D., {Lupsasca}, A., {Strominger}, A., {et~al.} 2020, Science Advances, 6, eaaz1310

\bibitem[{{Kocherlakota} {et~al.}(2024){Kocherlakota}, {Rezzolla}, {Roy}, \& {Wielgus}}]{Prashant2024}
{Kocherlakota}, P., {Rezzolla}, L., {Roy}, R., \& {Wielgus}, M. 2024, \mnras, 531, 3606

\bibitem[{Koposov {et~al.}(2023)Koposov, Speagle, Barbary, Ashton, Bennett, Buchner, Scheffler, Cook, Talbot, Guillochon, Cubillos, Ramos, Johnson, Lang, Ilya, Dartiailh, Nitz, McCluskey, Archibald, Deil, Foreman-Mackey, Goldstein, Tollerud, Leja, Kirk, Pitkin, Sheehan, Cargile, Patel, \& Angus}]{sergey_koposov_2023_7995596}
Koposov, S., Speagle, J., Barbary, K., {et~al.} 2023, Zenodo

\bibitem[{{Levis} {et~al.}(2024){Levis}, {Chael}, {Bouman}, {Wielgus}, \& {Srinivasan}}]{Aviad2024}
{Levis}, A., {Chael}, A.~A., {Bouman}, K.~L., {Wielgus}, M., \& {Srinivasan}, P.~P. 2024, Nature Astronomy, 8, 765

\bibitem[{{Lupsasca} {et~al.}(2024){Lupsasca}, {C{\'a}rdenas-Avenda{\~n}o}, {Palumbo}, {Johnson}, {Gralla}, {Marrone}, {Galison}, {Tiede}, \& {Keeble}}]{BHEX2024a}
{Lupsasca}, A., {C{\'a}rdenas-Avenda{\~n}o}, A., {Palumbo}, D. C.~M., {et~al.} 2024, arXiv e-prints, arXiv:2406.09498

\bibitem[{{Mo{\'s}cibrodzka} \& {Gammie}(2018)}]{Monika2018}
{Mo{\'s}cibrodzka}, M. \& {Gammie}, C.~F. 2018, \mnras, 475, 43

\bibitem[{{Najafi-Ziyazi} {et~al.}(2024){Najafi-Ziyazi}, {Davelaar}, {Mizuno}, \& {Porth}}]{Najafi2024}
{Najafi-Ziyazi}, M., {Davelaar}, J., {Mizuno}, Y., \& {Porth}, O. 2024, \mnras, 531, 3961

\bibitem[{{Narayan} {et~al.}(2003){Narayan}, {Igumenshchev}, \& {Abramowicz}}]{Narayan2003}
{Narayan}, R., {Igumenshchev}, I.~V., \& {Abramowicz}, M.~A. 2003, \pasj, 55, L69

\bibitem[{{Porth} {et~al.}(2021){Porth}, {Mizuno}, {Younsi}, \& {Fromm}}]{porth21}
{Porth}, O., {Mizuno}, Y., {Younsi}, Z., \& {Fromm}, C.~M. 2021, \mnras, 502, 2023

\bibitem[{Ripperda {et~al.}(2020)Ripperda, Bacchini, \& Philippov}]{Rip_2020_plasmoid}
Ripperda, B., Bacchini, F., \& Philippov, A.~A. 2020, The Astrophysical Journal, 900, 100

\bibitem[{{Ripperda} {et~al.}(2022){Ripperda}, {Liska}, {Chatterjee}, {Musoke}, {Philippov}, {Markoff}, {Tchekhovskoy}, \& {Younsi}}]{Ripperda2022}
{Ripperda}, B., {Liska}, M., {Chatterjee}, K., {et~al.} 2022, \apjl, 924, L32

\bibitem[{{Scepi} {et~al.}(2022){Scepi}, {Dexter}, \& {Begelman}}]{Scepi2022}
{Scepi}, N., {Dexter}, J., \& {Begelman}, M.~C. 2022, \mnras, 511, 3536

\bibitem[{{Speagle}(2020)}]{Speagle:2019ivv}
{Speagle}, J.~S. 2020, \mnras, 493, 3132

\bibitem[{{Tiede} {et~al.}(2020){Tiede}, {Pu}, {Broderick}, {Gold}, {Karami}, \& {Preciado-L{\'o}pez}}]{Tiede2020}
{Tiede}, P., {Pu}, H.-Y., {Broderick}, A.~E., {et~al.} 2020, \apj, 892, 132

\bibitem[{{Trippe} {et~al.}(2007){Trippe}, {Paumard}, {Ott}, {Gillessen}, {Eisenhauer}, {Martins}, \& {Genzel}}]{Trippe2007}
{Trippe}, S., {Paumard}, T., {Ott}, T., {et~al.} 2007, \mnras, 375, 764

\bibitem[{{Vincent} {et~al.}(2023){Vincent}, {Wielgus}, {Aimar}, {Paumard}, \& {Perrin}}]{Vincent2023}
{Vincent}, F.~H., {Wielgus}, M., {Aimar}, N., {Paumard}, T., \& {Perrin}, G. 2023, arXiv e-prints, arXiv:2309.10053

\bibitem[{{Vos} {et~al.}(2022){Vos}, {Mo{\'s}cibrodzka}, \& {Wielgus}}]{Vos:2022}
{Vos}, J., {Mo{\'s}cibrodzka}, M.~A., \& {Wielgus}, M. 2022, \aap, 668, A185

\bibitem[{{Vos} {et~al.}(2024){Vos}, {Olivares}, {Cerutti}, \& {Mo{\'s}cibrodzka}}]{Vos_2024_plasmoid}
{Vos}, J.~T., {Olivares}, H., {Cerutti}, B., \& {Mo{\'s}cibrodzka}, M. 2024, \mnras, 531, 1554

\bibitem[{{Walia} {et~al.}(2024){Walia}, {Kocherlakota}, {Chang}, \& {Salehi}}]{Rahul_2024aXv}
{Walia}, R.~K., {Kocherlakota}, P., {Chang}, D.~O., \& {Salehi}, K. 2024, arXiv e-prints, arXiv:2411.15119

\bibitem[{{Wielgus} {et~al.}(2022{\natexlab{a}}){Wielgus}, {Marchili}, {Mart{\'\i}-Vidal}, {Keating}, {Ramakrishnan}, {Tiede}, {Fomalont}, {Issaoun}, {Neilsen}, {Nowak}, {Blackburn}, {Gammie}, {Goddi}, {Haggard}, {Lee}, {Moscibrodzka}, {Tetarenko}, {Bower}, {Chan}, {Chatterjee}, {Chesler}, {Dexter}, {Doeleman}, {Georgiev}, {Gurwell}, {Johnson}, {Marrone}, {Mus}, {Psaltis}, {Ripperda}, {Witzel}, {Akiyama}, {Alberdi}, {Alef}, {Algaba}, {Anantua}, {Asada}, {Azulay}, {Bach}, {Baczko}, {Ball}, {Balokovi{\'c}}, {Barrett}, {Baub{\"o}ck}, {Benson}, {Bintley}, {Blundell}, {Boland}, {Bouman}, {Boyce}, {Bremer}, {Brinkerink}, {Brissenden}, {Britzen}, {Broderick}, {Broguiere}, {Bronzwaer}, {Bustamante}, {Byun}, {Carlstrom}, {Ceccobello}, {Chael}, {Chatterjee}, {Chen}, {Chen}, {Cho}, {Christian}, {Conroy}, {Conway}, {Cordes}, {Crawford}, {Crew}, {Cruz-Osorio}, {Cui}, {Davelaar}, {De Laurentis}, {Deane}, {Dempsey}, {Desvignes}, {Dhruv}, {Dzib}, {Eatough}, {Emami}, {Falcke}, {Farah}, {Fish}, {Ford}, {Fraga-Encinas},
  {Freeman}, {Friberg}, {Fromm}, {Fuentes}, {Galison}, {Garc{\'\i}a}, {Gentaz}, {Gold}, {G{\'o}mez-Ruiz}, {G{\'o}mez}, {Gu}, {Hada}, {Haworth}, {Hecht}, {Hesper}, {Ho}, {Ho}, {Honma}, {Huang}, {Huang}, {Hughes}, {Ikeda}, {Impellizzeri}, {Inoue}, {James}, {Jannuzi}, {Janssen}, {Jeter}, {Jiang}, {Jim{\'e}nez-Rosales}, {Jorstad}, {Joshi}, {Jung}, {Karami}, {Karuppusamy}, {Kawashima}, {Kettenis}, {Kim}, {Kim}, {Kim}, {Kim}, {Kino}, {Koay}, {Kocherlakota}, {Kofuji}, {Koch}, {Koyama}, {Kramer}, {Kramer}, {Krichbaum}, {Kuo}, {La Bella}, {Lauer}, {Lee}, {Leung}, {Levis}, {Li}, {Lico}, {Lindahl}, {Lindqvist}, {Lisakov}, {Liu}, {Liu}, {Liuzzo}, {Lo}, {Lobanov}, {Loinard}, {Lonsdale}, {Lu}, {Mao}, {Markoff}, {Marscher}, {Matsushita}, {Matthews}, {Medeiros}, {Menten}, {Michalik}, {Mizuno}, {Mizuno}, {Moran}, {Moriyama}, {M{\"u}ller}, {Musoke}, {Myserlis}, {Nadolski}, {Nagai}, {Nagar}, {Nakamura}, {Narayan}, {Narayanan}, {Natarajan}, {Nathanail}, {Navarro Fuentes}, {Neri}, {Ni}, {Noutsos}, {Oh}, {Okino}, {Olivares},
  {Ortiz-Le{\'o}n}, {Oyama}, {{\"O}zel}, {Palumbo}, {Paraschos}, {Park}, {Parsons}, {Patel}, {Pen}, {Pesce}, {Pi{\'e}tu}, {Plambeck}, {PopStefanija}, {Porth}, {P{\"o}tzl}, {Prather}, {Preciado-L{\'o}pez}, {Pu}, {Rao}, {Rawlings}, {Raymond}, {Rezzolla}, {Ricarte}, {Roelofs}, {Rogers}, {Ros}, {Romero-Canizales}, {Roshanineshat}, {Rottmann}, {Roy}, {Ruiz}, {Ruszczyk}, {Rygl}, {S{\'a}nchez}, {S{\'a}nchez-Arg{\"u}elles}, {S{\'a}nchez-Portal}, {Sasada}, {Satapathy}, {Savolainen}, {Schloerb}, {Schuster}, {Shao}, {Shen}, {Small}, {Won Sohn}, {SooHoo}, {Souccar}, {Sun}, {Tazaki}, {Tilanus}, {Titus}, {Torne}, {Traianou}, {Trent}, {Trippe}, {van Bemmel}, {van Langevelde}, {van Rossum}, {Vos}, {Wagner}, {Ward-Thompson}, {Wardle}, {Weintroub}, {Wex}, {Wharton}, {Wiik}, {Wondrak}, {Wong}, {Wu}, {Yamaguchi}, {Yoon}, {Young}, {Young}, {Younsi}, {Yuan}, {Yuan}, {Zensus}, {Zhang}, {Zhao}, \& {Zhao}}]{Wielgus2022_LC}
{Wielgus}, M., {Marchili}, N., {Mart{\'\i}-Vidal}, I., {et~al.} 2022{\natexlab{a}}, \apjl, 930, L19

\bibitem[{{Wielgus} {et~al.}(2022{\natexlab{b}}){Wielgus}, {Moscibrodzka}, {Vos}, {Gelles}, {Mart{\'\i}-Vidal}, {Farah}, {Marchili}, {Goddi}, \& {Messias}}]{W22}
{Wielgus}, M., {Moscibrodzka}, M., {Vos}, J., {et~al.} 2022{\natexlab{b}}, \aap, 665, L6, (W22)

\bibitem[{{Wong}(2021)}]{Wong2021}
{Wong}, G.~N. 2021, \apj, 909, 217

\bibitem[{{Yfantis} {et~al.}(2024{\natexlab{a}}){Yfantis}, {Wielgus}, \& {Moscibrodzka}}]{yfantis24b}
{Yfantis}, A., {Wielgus}, M., \& {Moscibrodzka}. 2024{\natexlab{a}}, \aap, 691, A327

\bibitem[{{Yfantis} {et~al.}(2024{\natexlab{b}}){Yfantis}, {Mo{\'s}cibrodzka}, {Wielgus}, {Vos}, \& {Jimenez-Rosales}}]{yfantis24a}
{Yfantis}, A.~I., {Mo{\'s}cibrodzka}, M.~A., {Wielgus}, M., {Vos}, J.~T., \& {Jimenez-Rosales}, A. 2024{\natexlab{b}}, \aap, 685, A142

\end{thebibliography}

\end{document}